\documentclass[aps,floatfix,eqsecnum,showpacs,preprint,tightenlines]{revtex4-1}
\usepackage{epsfig}
\usepackage{bm}

\usepackage{amsmath}

\def\e{\kern+.5ex\lower.42ex\hbox{$\scriptstyle \iota$}\kern-1.10ex e}
\def\registered{{\ooalign{\hfil\raise .00ex\hbox{\scriptsize R}\hfil\crcr\mathhexbox20D}}}

\newcommand{\BA}[1]{\langle #1 \mid}

\newcommand{\KT}[1]{\mid #1 \rangle}

\begin{document}


\title{Break-up channels in muon capture on $^3$He}



\author{J. Golak}
\affiliation{M. Smoluchowski Institute of Physics, Jagiellonian University, PL-30059 Krak\'ow, Poland}
\author{R. Skibi{\'n}ski}
\affiliation{M. Smoluchowski Institute of Physics, Jagiellonian University, PL-30059 Krak\'ow, Poland}
\author{H. Wita{\l}a}
\affiliation{M. Smoluchowski Institute of Physics, Jagiellonian University, PL-30059 Krak\'ow, Poland}
\author{K. Topolnicki}
\affiliation{M. Smoluchowski Institute of Physics, Jagiellonian University, PL-30059 Krak\'ow, Poland}
\author{A.E. Elmeshneb}
\affiliation{M. Smoluchowski Institute of Physics, Jagiellonian University, PL-30059 Krak\'ow, Poland}
\author{H. Kamada}
\affiliation{Department of Physics, Faculty of Engineering,
Kyushu Institute of Technology, Kitakyushu 804-8550, Japan}
\author{A. Nogga}
\affiliation{Forschungszentrum J\"ulich,
          Institut f\"ur Kernphysik (Theorie),
          Institute for Advanced Simulation
          and J\"ulich Center for Hadron Physics, D-52425 J\"ulich, Germany}
\author{L.E. Marcucci}
\affiliation{Department of Physics, University of Pisa, IT-56127 Pisa, Italy
              and INFN-Pisa, IT-56127 Pisa, Italy}


\date{\today}

\begin{abstract}
The
$\mu^- + ^2{\rm H} \rightarrow \nu_\mu + n + n$, 
$\mu^- + ^3{\rm He} \rightarrow \nu_\mu + ^3{\rm H}$, 
$\mu^- + ^3{\rm He} \rightarrow \nu_\mu + n + d$ 
and 
$\mu^- + ^3{\rm He} \rightarrow \nu_\mu + n + n + p$
capture reactions
are studied with various realistic potentials
under full inclusion of final state interactions.
Our results for the two- and three-body break-up of $^3$He 
are calculated with a variety of nucleon-nucleon potentials, among which is
the AV18 potential, augmented by the Urbana~IX 
three-nucleon potential. Most of our results are based on 
the single nucleon weak current operator. As a first step, we have
tested our calculation
in the case of the 
$\mu^- + ^2{\rm H} \rightarrow \nu_\mu + n + n$
and 
$\mu^- + ^3{\rm He} \rightarrow \nu_\mu + ^3{\rm H}$ reactions, for which
theoretical predictions obtained in a comparable framework
are available.
Additionally, we have been able to obtain for the first time 
a realistic estimate for the total rates of the muon capture reactions
on $^3$He in the break-up channels: 544 s$^{-1}$ 
and 154 s$^{-1}$ for the $n + d$ and 
$n + n + p$ channels, respectively.
Our results have also been compared with the most recent experimental
data, finding a rough agreement for the total capture rates,
but failing to reproduce the differential capture rates.
\end{abstract}


\maketitle


\section{Introduction}
\label{section1}

Muon capture reactions on light nuclei have been studied intensively both experimentally 
and theoretically for many years. For informations on earlier achievements we refer the reader to Refs.~\cite{Mea01,Gor04,Kam10}. 
More recent theoretical work, focused
on the $ \mu^- + ^2{\rm H} \rightarrow \nu_\mu + n + n $
and
$ \mu^- + ^3{\rm He} \rightarrow \nu_\mu + ^3{\rm H} $
reactions, has been summarized in Refs.~\cite{prc83.014002,Mar12}.
Here we mention only that 
the calculation of Ref.~\cite{prc83.014002}, following the early steps
of Ref.~\cite{Mar02}, was performed both in the phenomenological and the
``hybrid'' chiral effective field theory ($\chi$EFT) approach. In the first one,
Hamiltonians based
on conventional two-nucleon (2N)
and three-nucleon (3N) potentials were used to calculate the nuclear wave functions,
and the weak transition operator included, beyond the single nucleon 
contribution
associated with the basic process $\mu^- +p \rightarrow \nu_\mu + n$,
meson-exchange currents as well as
currents arising from the excitation of $\Delta$-isobar degrees of 
freedom~\cite{prc63.015801}.
In  the hybrid $\chi$EFT approach,
the weak operators were derived in $\chi$EFT, but
their matrix elements were evaluated between wave functions
obtained from conventional potentials.  Typically, the potential model 
and
hybrid $\chi$EFT predictions are in good agreement with each 
other~\cite{prc83.014002}.
Only very recently, the two reactions have been studied in a ``non-hybrid''
$\chi$EFT approach~\cite{Mar12b}, where both potentials and currents are
derived consistently in $\chi$EFT and the low-energy constants 
present in the
3N potential and two-body axial-vector current are constrained
to reproduce the $A=3$ binding energies and the 
Gamow-Teller matrix element in tritium $\beta$-decay. 
An overall agreement between the results
obtained within different approaches has been found, as well as between
theoretical predictions and available experimental data.

The first theoretical study for the capture
$\mu^- + ^3{\rm He} \rightarrow \nu_\mu + n + d$ 
was reported in Ref.~\cite{romek}.
A simple single nucleon current operator was used
without any relativistic corrections
and the initial and final 3N states
were generated using realistic nucleon-nucleon potentials but neglecting the 
3N interactions. 

Recent progress in few-nucleon calculations has prompted us to 
join our expertises: from momentum space treatment of 
electromagnetic processes \cite{physrep,romek2} 
and by using the potential model approach developed in 
Ref.~\cite{prc83.014002}. We neglect as a first step meson-exchange
currents and perform a systematic study of all the $A=2$ and $A=3$
muon capture reactions, extending the calculations of Ref.~\cite{romek} 
to cover also the $\mu^- + ^3{\rm He} \rightarrow \nu_\mu + n + n + p$
channel. 
Therefore, the motivation behind this work is twofold:
first of all, by comparing our results obtained for the 
$ \mu^- + ^2{\rm H} \rightarrow \nu_\mu + n + n $
and
$ \mu^- + ^3{\rm He} \rightarrow \nu_\mu + ^3{\rm H} $
reactions with those of Ref.~\cite{prc83.014002}, we will be able to 
establish a theoretical framework which can be extended to all 
the $A\leq 3$ muon capture reactions, including those which involve the full
break-up of the $A=3$ final state. Note that the results of 
Ref.~\cite{prc83.014002} were obtained using the hyperspherical harmonics
formalism (for a review, see Ref.~\cite{Kie08}), at present not available for the
$A=3$ full break-up channel. Here, by using the Faddeev equation
approach, this difficulty is overcome.

The second motivation behind this work is that we will provide, for the 
first time, predictions for the total and differential capture rates of the
reactions
$\mu^- + ^3{\rm He} \rightarrow \nu_\mu + n + d$ 
and 
$\mu^- + ^3{\rm He} \rightarrow \nu_\mu + n + n + p$,
obtained with full inclusion of final state interactions, not only
nucleon-nucleon but also 3N forces.

The paper is organized in the following way.
In Sec.~\ref{section2} we introduce the single nucleon 
current operator, which we treat exclusively in momentum space, and
compare our expressions with those of Ref.~\cite{prc83.014002}.
In the following two sections we show selected results 
for the $\mu^- + ^2{\rm H} \rightarrow \nu_\mu + n + n$ (Sec.~\ref{section3})
and
for the $\mu^- + ^3{\rm He} \rightarrow \nu_\mu + ^3{\rm H}$ (Sec.~\ref{section4})
reactions.
Since these results are obtained by retaining only the single nucleon
current operator, a comparison with those of Ref.~\cite{prc83.014002},
where meson-exchange currents were included,
will inform the reader about the theoretical error caused by neglecting 
all contributions beyond the single nucleon term.

Our main results are shown in Sec.~\ref{section5},
where we discuss in detail the way we calculate the total 
capture rates for the two break-up reactions,
$\mu^- + ^3{\rm He} \rightarrow \nu_\mu + n + d$
and
$\mu^- + ^3{\rm He} \rightarrow \nu_\mu + n + n + p$,
and show predictions obtained with different 3N 
dynamics. 
In these calculations we
employ mainly the AV18 nucleon-nucleon potential~\cite{av18}
supplemented with the Urbana~IX 3N potential~\cite{urbana}.
These results form a solid base for our future calculations where the
meson-exchange currents will be included, and
provide a set of benchmark results.
Note that in Secs.~\ref{section6} and \ref{section6.5} 
we provide an analysis of the most recent 
(from Ref.~\cite{pra69.012712}) 
and the older (from Refs.~\cite{datainromek1,datainromek2})
experimental data 
on differential capture rates 
for the reactions
$\mu^- + ^3{\rm He} \rightarrow \nu_\mu + n + d$
and
$\mu^- + ^3{\rm He} \rightarrow \nu_\mu + n + n + p$. Finally,
Sec.~\ref{section7} contains some concluding remarks.

\section{The single nucleon current operator}
\label{section2}

In the muon capture process we assume that the initial state 
$ \KT{i\, } $ consists of the atomic $K$-shell muon wave function 
$ \KT{ \psi \, m_\mu \, } $ with the muon spin projection $m_\mu$ 
and the initial nucleus state with the three-momentum ${\bf P}_i$ 
(and the spin
projection $m_{i}$):
\begin{eqnarray}
\KT{i\, } = \KT{\psi \, m_\mu \, } \, \KT{\Psi_i \, {\bf P}_i \, m_{i} \, } \, .
\label{i}
\end{eqnarray}
In the final state, $\KT{f\, }$,  one encounters
the muon neutrino 
(with the three-momentum ${\bf p}_{\nu}$ 
and the spin projection $m_{\nu}$), 
as well as the final nuclear state with the total 
three-momentum ${\bf P}_f$ and the set of spin projections $m_{f}$:
\begin{eqnarray}
\KT{f\, } = \KT{\nu_\mu \, {\bf p}_{\nu} \, m_\nu \, } \, 
\KT{\Psi_f \, {\bf P}_f \, m_{f} \, } \, .
\label{f}
\end{eqnarray}
The transition from the initial to final state is driven by the 
Fermi form of the interaction Lagrangian (see for example Ref.~\cite{walecka})
and leads to a contraction of the leptonic (${\cal L}_\lambda$) and nuclear 
(${\cal N}^\lambda$) parts 
in the $S$-matrix element, $S_{fi}$ \cite{romek}:
\begin{eqnarray}
S_{fi}= i ( 2 \pi )^4 \, \delta^4 \left( P^\prime - P \right)\, 
\frac {G}{\sqrt{2}} \, {\cal L}_\lambda \, {\cal N}^\lambda \, ,
\label{sfi}
\end{eqnarray}
where $G= 1.14939 \times 10^{-5} \, {\rm GeV}^{-2} $ is the Fermi constant
(taken from Ref.~\cite{prc83.014002}), 
and $P$ ($P^\prime$) is the 
total initial (final) four-momentum. 
The well known leptonic matrix element 
\begin{eqnarray}
{\cal L}_\lambda = \frac 1{ \left( 2 \pi \, \right)^3 } \, 
\bar{u} ( {\bf p}_{\nu} , m_{\nu} ) \gamma_\lambda ( 1- \gamma_5 ) 
u ( {\bf p}_{\mu} , m_{\mu} )
\, \equiv \,
\frac 1{ \left( 2 \pi \, \right)^3 } \, L_\lambda
\label{llambda}
\end{eqnarray}
is given in terms of the Dirac spinors (note that we use 
the notation and spinor normalization of Bjorken and Drell \cite{bjodrell}).

The nuclear part is the essential ingredient of the formalism,
and is written as
\begin{eqnarray}
{\cal N}^\lambda = \frac 1{ \left( 2 \pi \, \right)^3 } \, 
\BA{\Psi_f \, {\bf P}_f \, m_{f} \, } \, 
j_w^\lambda
\, \KT{\Psi_i \, {\bf P}_i \, m_{i} \, } 
\, \equiv \,
\frac 1{ \left( 2 \pi \, \right)^3 } \, N^\lambda \, .
\label{nlambda}
\end{eqnarray}
It is a matrix element of the nuclear weak current operator 
$j_w^\lambda$ between the initial and final nuclear states. 
The primary form of $N^\lambda $ is present already in such basic 
processes (from the point of view of the Fermi theory) 
as the neutron beta decay or the low-energy 
$\mu^- + p \rightarrow \nu_\mu + n $ reaction.
General considerations, taking into account symmetry 
requirements, lead to the following form of the single nucleon 
current operator \cite{bailin82},
whose matrix elements depend on the nucleon incoming 
(${\bf p}$) 
and outgoing momentum 
($ {\bf p}^{\, \prime}$)
and nucleon spin projections $m$  and $m'$:
\begin{eqnarray}
&&\BA{ \frac12 m'} 
\BA{{\bf p}^{\, \prime}} 
{j}_w^{\lambda}(1) 
\KT{{\bf p}} 
\KT{ \frac12 m} 
= \nonumber \\
&& \bar{{u}}({\bf p}^{\, \prime} , m') \Big( \left(g_{1}^{V} - 2 M \, g_{2}^{V}\right) \gamma^{\lambda} + g_{2}^{V} 
 \left(p + p'\, \right)^{\lambda}  \nonumber \\
&&+ g_{1}^{A} \gamma^{\lambda} \gamma^{5} + g_{2}^{A} \left(p - p' \, \right)^{\lambda} \gamma^{5} \Big) 
 {\tau}_{-} {u}({\bf p} , m) \, ,
\label{j1rel}
\end{eqnarray}
containing nucleon weak form factors,
$g_{1}^{V}$,
$g_{2}^{V}$,
$g_{1}^{A}$, and $g_{2}^{A}$, which are functions 
of the four-momentum transfer squared, $ ( p ' - p )^2$.
We neglect the small difference between the proton mass $M_p$ 
and neutron mass $M_n$
and introduce the average ``nucleon mass'', 
$M \equiv \frac12 \left( M_p + M_n \, \right) $.
Working with the isospin formalism, we introduce the
isospin lowering operator, as 
${\tau}_{-}=(\tau_x -{\rm i} \tau_y)/2$.
Since the wave functions are generated by nonrelativistic
equations, it is necessary to perform the nonrelativistic reduction 
of Eq.~(\ref{j1rel}). The nonrelativistic form of the time
and space components of $j_w^\lambda (1)$ reads 
\begin{eqnarray}
\BA{{\bf p}^{\, \prime}} {j}^{0}_{\text{NR}}(1)  \KT{{\bf p}} = \left( g_{1}^{V} + g_{1}^{A} \frac{{{\bm \sigma}} 
\cdot \left( {\bf p} + {\bf p}^{\, \prime} \right)}{2 M}  \right) {\tau}_{-}
\label{jnr01}
\end{eqnarray}
and
\begin{eqnarray}
&&\BA{{\bf p}^{\, \prime}} {{\bf j}}_{\text{NR}}(1)  \KT{{\bf p}\, } 
= \nonumber \\
&&\bigg( g_{1}^{V} \frac{{\bf p} + {\bf p}^{\, \prime}}{2 M} - \frac{1}{2 M} \left(g_{1}^{V} - 2 M g_{2}^{V} \, \right) 
i \, {{\bm \sigma}} \times \left( {\bf p} - {\bf p}^{\, \prime}  \, \right) \nonumber \\
&&+ g_{1}^{A} {{\bm \sigma}} + g_{2}^{A} \left( {\bf p} - {\bf p}^{\, \prime} \, \right) \frac{{{\bm \sigma}} \cdot 
\left({\bf p} - {\bf p}^{\, \prime} \, \right)}{2 M} \bigg) {\tau}_{-} \, ,
\label{jnrvec1}
\end{eqnarray}
where ${{\bm \sigma}}$ is a vector of Pauli spin operators. Here we
have kept only terms up to $1/M$.

Very often relativistic $1/M^2$ corrections are also included. 
This leads then to additional terms in the current operator:  
\begin{eqnarray}
&&\BA{{\bf p}^{\, \prime}} {j}^{0}_{\text{NR+RC}}(1)  \KT{{\bf p}} = \nonumber \\
&&\bigg( g_{1}^{V} - (g_{1}^{V} - 4 M g_{2}^{V}) \frac{\left({\bf p}^{\, \prime} - {\bf p} \, \right)^{2}}{8 M^{2}} + 
\left(g_{1}^{V} - 4 M g_{2}^{V} \, \right) \, i \, \frac{\left( {\bf p}^{\, \prime} \times {\bf p} \, \right) \cdot 
{{\bm \sigma}}}{4 M^{2}} \nonumber \\
&&+ g_{1}^{A} \frac{{{\bm \sigma}} \cdot \left( {\bf p} + {\bf p}^{\, \prime} \, \right)}{2 M} + g_{2}^{A} 
\frac{\left( {\bf p}^{\, \prime\, 2} - {\bf p}^{2} \, \right)}{4 M^{2}} {{\bm \sigma}} \cdot \left({\bf p}^{\, \prime} - 
{\bf p} \, \right)  \bigg) {\tau}_{-} 
\label{1nr+rc01}
\end{eqnarray}
and
\begin{eqnarray}
&&\BA{{\bf p}^{\, \prime}} {{\bf j}}_{\text{NR+RC}}(1)  \KT{{\bf p}} = \nonumber \\
&&\bigg( g_{1}^{V} \frac{{\bf p} + {\bf p^{\, \prime}}}{2 M} - \frac{1}{2 M} \left(g_{1}^{V} - 2 M g_{2}^{V} \, \right) 
\, i \, {{\bm \sigma}} \times \left( {\bf p} - {\bf p}^{\, \prime} \, \right) \nonumber \\
&&+ g_{1}^{A} \left(1 - \frac{\left({\bf p} + {\bf p}^{\, \prime} \, \right)^{2}}{8 M^{2}} \, \right) {{\bm \sigma}} + 
\nonumber \\
&&+ \frac{g_{1}^{A}}{4 M^{2}} \big[ \left({\bf p} \cdot {{\bm \sigma}} \, \right) {\bf p}^{\, \prime} + \left({\bf p}^{\, \prime} 
\cdot {{\bm \sigma}} \, \right) {\bf p} + \, i \, \left( {\bf p} \times 
{\bf p}^{\, \prime} \,  \right) \big]) \nonumber \\
&&+g_{2}^{A} \left( {\bf p} - {\bf p}^{\, \prime} \, \right) \frac{{{\bm \sigma}} \cdot \left({\bf p} - {\bf p}^{\, \prime} \, \right)}{2 M} 
\bigg) {\tau}_{-} \ .
\label{1nr+rcvec1}
\end{eqnarray}
This form of the nuclear weak current operator is 
very close to the one used in Ref.~\cite{prc83.014002}, provided that one term,
\begin{eqnarray}
g_{2}^{V} \frac{\left({\bf p}^{\, \prime} - {\bf p} \, \right)^{2}}{2 M} 
\label{extraterm}
\end{eqnarray}
is dropped in Eq.~(\ref{1nr+rc01}) and 
we use:
\begin{eqnarray}
G_E^V&=&g_1^V  \, , \label{gev} \\
G_M^V&=&g_1^V-2M g_2^V \, , \label{gmv} \\
G_{A} &=& -g_{1}^{A} \, , \label{ga}\\
G_{P} &=& -g_{2}^{A} m_{\mu} \, . \label{gp}
\end{eqnarray}
Here
the form factors 
$G_E^V$ and $G_M^V$ are the isovector components of the electric and magnetic
Sachs form factors, while $G_{A}$ and $G_{P}$ are the axial and pseudoscalar 
form factors. Their explicit expressions and parametrization can be found
in Ref.~\cite{She12}.
We also verified that the extra term (\ref{extraterm}) 
gives negligible effects in all studied observables. 

It is clear that on top of the single nucleon operators,
also many-nucleon contributions appear in $j_w^\lambda$.
In the 3N system one can even expect 3N 
current operators:
\begin{eqnarray}
j_w^\lambda = 
j_w^\lambda (1) + 
j_w^\lambda (2) + 
j_w^\lambda (3) + 
j_w^\lambda (1,2) + 
j_w^\lambda (1,3) + 
j_w^\lambda (2,3) + 
j_w^\lambda (1,2,3)
\label{jwmany} \, .
\end{eqnarray}
The role of these many-nucleon operators has been studied 
for example in Ref.~\cite{prc83.014002}. 
In spite of the progress made in this direction 
(see the discussion in Ref.~\cite{prc83.014002}), 
we decided 
to base our first predictions on the single nucleon current 
only and concentrate on other dynamical ingredients. 
Since we want to compare our results with the ones published in 
Ref.~\cite{prc83.014002}, we start with the 
$\mu^- + ^2{\rm H} \rightarrow \nu_\mu + n + n $ 
and
$\mu^- + ^3{\rm He}  \rightarrow \nu_\mu + ^3{\rm H}$ 
reactions. Although the steps leading from the general form of $S_{fi}$ 
to the capture rates formula are standard, we give here formulas for kinematics 
and
capture rates for all the studied reactions, expecting that they might 
become useful in future benchmark calculations.

\section{Results for the $\mu^- + ^2{\rm H} \rightarrow \nu_\mu + n + n $ reaction}
\label{section3}

The kinematics of this processes can be treated 
without any approximations both relativistically and nonrelativistically.
We make sure that the nonrelativistic approximation is fully justified 
by comparing values of various quantities calculated nonrelativistically 
and using relativistic equations. This is important, since our dynamics 
is entirely nonrelativistic.
In all cases the starting point is the energy and momentum conservation,
where we neglect the very small binding energy of the muon atom and 
the neutrino mass, assuming that the initial deuteron and muon are at rest.
In the case of the $\mu^- + ^2{\rm H} \rightarrow \nu_\mu + n + n $ reaction 
it reads
\begin{eqnarray} 
M_\mu + M_d &=&  E_\nu +
\sqrt{ M_n^2 + {\bf p}_1^{\ 2} \, }   
+
\sqrt{ M_n^2 + {\bf p}_2^{\ 2} \, }  \, , \nonumber  \\
{\bf p}_1 +  {\bf p}_2 +  {\bf p}_\nu &=&  0 
\label{relnn}
\end{eqnarray}  
and the first equation in (\ref{relnn}) is approximated nonrelativistically by 
\begin{eqnarray} 
M_\mu + M_d =  E_\nu +
2 M_n + 
\frac { {\bf p}_1^{\ 2} \, }  { 2 M_n}  
+
\frac { {\bf p}_2^{\ 2} \, }  { 2 M_n}  \, .
\label{nrlnn}
\end{eqnarray}  
The maximal relativistic and non-relativistic 
neutrino energies read correspondingly
\begin{eqnarray} 
\left( E_\nu^{max,nn} \, \right)^{rel} =
\frac{1}{2} \left(-\frac{4 {M_n}^2}{{M_d}+{M_\mu}}+ {M_d}+{M_\mu}\right)
\end{eqnarray}  
and
\begin{eqnarray} 
\left( E_\nu^{max,nn} \, \right)^{nrl} =
2 \sqrt{{M_d} {M_n}+{M_\mu}
   {M_n}-{M_n}^2}-2 {M_n} \, .
\end{eqnarray}  
Assuming 
$ M_p$ = 938.272 MeV,
$ M_n$ = 939.565 MeV,
$M_\mu$ = 105.658 MeV,
$M_d = M_p + M_n$ - 2.225 MeV, we obtain
$ \left( E_\nu^{max,nn} \, \right)^{rel} $ = 99.5072 MeV
and
$ \left( E_\nu^{max,nn} \, \right)^{nrl} $ = 99.5054 MeV, respectively,
with a difference which is clearly negligible.

Further we introduce the relative Jacobi momentum,
$ {\bf p} = \frac12 \,  \left( {\bf p}_1 -  {\bf p}_2 \, \right) $,
and write the energy conservation in a way which best corresponds to the
nuclear matrix element calculations:
\begin{eqnarray} 
M_\mu + M_d =  E_\nu +
2 M_n + 
\frac { E_\nu^{2} \, }  { 4 M_n}  
+
\frac { {\bf p}^{\ 2} \, }  { M_n}  \, .
\label{nrlnn2}
\end{eqnarray}  

In the nuclear matrix element,
$\BA{\Psi_f \, {\bf P}_f \, m_{f} \, } \, 
j_w^\lambda \, \KT{\Psi_i \, {\bf P}_i \, m_{i} \, } $,
we deal with the deuteron in the initial state 
and with a two-neutron scattering state in the final state.
Introducing the spin magnetic quantum numbers, we write
\begin{eqnarray} 
\BA{\Psi_f \, {\bf P}_f \, m_{f} \, } \, 
j_w^\lambda \, \KT{\Psi_i \, {\bf P}_i \, m_{i} \, } \, = \,
^{(-)}\BA{ {\bf p} \  {\bf P}_f=-{\bf p}_\nu \, m_{1} \, m_2 } \, 
j_w^\lambda \, \KT{\phi_d \, {\bf P}_i=0 \, m_{d} \, } \, \nonumber \\
= \ \BA{ {\bf p} \  {\bf P}_f=-{\bf p}_\nu \, m_{1} \, m_2 } \, 
\Big( \,  1 +  t (E_{nn} ) \, G_0^{nn} (E_{nn} ) \, \Big) \, 
j_w^\lambda \, \KT{\phi_d \, {\bf P}_i=0 \, m_{d} \, } \, .
\label{nnn1}
\end{eqnarray}  
Thus for a given nucleon-nucleon potential, $V$, 
the scattering state of two neutrons is generated 
by introducing the solution of the Lippmann-Schwinger equation, $t$:
\begin{eqnarray} 
t (E_{nn}) = V +  t (E_{nn} ) \, G_0^{nn} (E_{nn} ) \, V \, ,
\label{t}
\end{eqnarray} 
where $ G_0^{nn} (E_{nn}) $ is the free 2N propagator 
and the relative energy in the two-neutron system 
is 
\begin{eqnarray}
E_{nn} = \frac { {\bf p}^{\, 2} \, }  { M_n}  =
M_\mu + M_d -  E_\nu - 2 M_n - \frac { E_\nu^{2} \, }  { 4 M_n} \, .
\label{enn}
\end{eqnarray}  
 
We generate the deuteron wave function and 
solve Eq.~(\ref{t}) in momentum space.
Note that here, as well as for the $A=3$ systems, we use the
avarage ``nucleon mass'' in the kinematics and in solving the
Lippmann-Schwinger equation. The effect of this approaximation on the
$\mu^-+^2{\rm H}\rightarrow \nu_\mu+n+n$ reaction will be discussed
below.
Taking all factors into account and evaluating the phase space factor 
in terms of the relative momentum, we arrive at the following 
expression for the total capture rate

\begin{eqnarray}
&&\Gamma_d = \frac12 G^2 \frac1{( 2 \pi )^2 } \, \frac { \left(  M^\prime_d \alpha \, \right)^3 } {\pi  } \, 
\int\limits_{0}^{\pi} d \theta_{p_\nu}  \sin \theta_{p_\nu}  \, \int\limits_{0}^{2 \pi} d \phi_{p_\nu}  \, 
\int\limits_0^{E_\nu^{max,nn}} \, dE_\nu E_\nu^2  \,
\frac12 M_n p  \,  
\nonumber \\
&&\int\limits_{0}^{\pi} d \theta_p \sin \theta_p \, \int\limits_{0}^{2 \pi} d \phi_p \, 
\frac16 \sum\limits_{m_d,m_\mu} \sum\limits_{m_1, m_2, m_\nu } 
\left| L_\lambda ( m_\nu , m_\mu \, ) N^\lambda (m_1, m_2, m_d \, ) \, \right|^2  \, ,
\label{gnn1}
\end{eqnarray}  
where 
the factor $ \frac { \left(  M^\prime_d \alpha \, \right)^3 } {\pi  } $ stems from the 
$K$-shell atomic wave function, $ M^\prime_d  = \frac { M_d M_\mu } { M_d + M_\mu }$ 
and  $ \alpha \approx \frac 1{137} $ is the fine structure constant. 
We can further simplify this expression, since for the unpolarized case the integrand 
does not depend on the neutrino direction and the azimuthal angle of the relative 
momentum, $\phi_p$. Thus we set $ {\hat {\bf p}}_\nu = - {\hat {\bf z}} $,
choose $\phi_p = 0$ and introduce the explicit
components of $ N^\lambda (m_1, m_2, m_d \, )$, which yields
\begin{eqnarray}
&&\Gamma_d = \frac12 G^2 \frac1{( 2 \pi )^2 } \, \frac { \left(  M^\prime_d \alpha \, \right)^3 } {\pi  } \, 
4 \pi \,
\int\limits_0^{E_\nu^{max,nn}} \, dE_\nu E_\nu^2  \,
\frac12 M p  \,  
\nonumber \\
&&2 \pi \, \int\limits_{0}^{\pi} d \theta_p \sin \theta_p \, 
\frac13 \sum\limits_{m_d} \sum\limits_{m_1, m_2 } 
\Big( 
\left| N^0 (m_1, m_2, m_d \, ) \, \right |^2 \, + \, 
\left| N_z (m_1, m_2, m_d \, ) \, \right |^2 \, + \, 
\nonumber \\
&&2 \left| N_{-1} (m_1, m_2, m_d \, ) \, \right |^2 \, + \, 
2 {\rm Re} \left( N^0 (m_1, m_2, m_d \, )  \left( N_z (m_1, m_2, m_d \, ) \right)^*   \right ) \, \Big) \,  .
\label{gnn2}
\end{eqnarray}  

This form is not appropriate when we want to calculate separately capture rates 
from two hyperfine states $F=\frac12$ or $ F=\frac32$ of the muon-deuteron atom.
In such a case we introduce the coupling between the deuteron and muon spin
via standard Clebsch-Gordan coefficients $c ( \frac12 , 1 , F ; m_\mu , m_d , m_F \, ) $ 
and obtain
\begin{eqnarray}
&&\Gamma_d^F = \frac12 G^2 \frac1{( 2 \pi )^2 } \, \frac { \left(  M^\prime_d \alpha \, \right)^3 } {\pi  } \, 
4 \pi \,
\int\limits_0^{E_\nu^{max,nn}} \, dE_\nu E_\nu^2  \,
\frac12 M p  \,  
\nonumber \\
&&2 \pi \, \int\limits_{0}^{\pi} d \theta_p \sin \theta_p \, 
\frac1{2F + 1} \sum\limits_{m_F} \sum\limits_{m_1, m_2 , m_\nu}  \nonumber \\
&&\Big| 
\sum\limits_{m_\mu , m_d} \, 
c ( \frac12 , 1 , F ; m_\mu , m_d , m_F \, ) \, 
L_\lambda ( m_\nu , m_\mu \, ) N^\lambda (m_1, m_2, m_d \, ) \, 
\Big|^2 \, .
\label{gnnf}
\end{eqnarray}  
For the sake of clarity, in Eqs.~(\ref{gnn1})--(\ref{gnnf})
we show the explicit dependence of $N^\lambda $ on the 
spin magnetic quantum numbers.

\begin{figure}
\includegraphics[width=7cm]{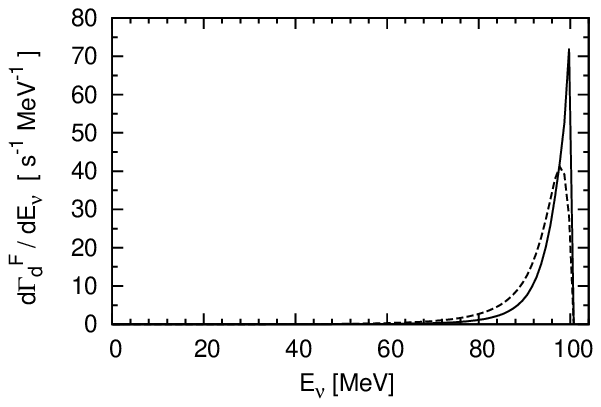}
\includegraphics[width=7cm]{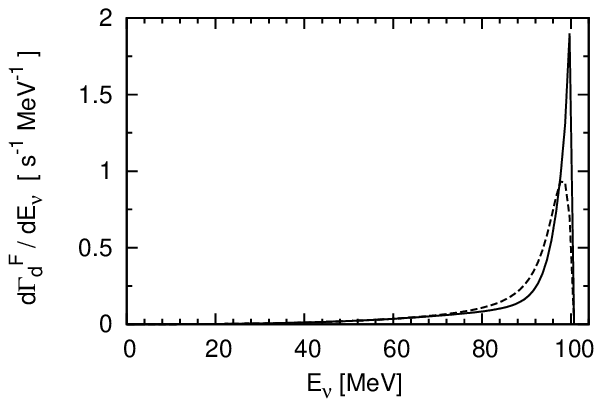}
\caption{Differential capture rate $ {d\Gamma_d^F }/ {dE_\nu} $
for the $\mu^-+ ^2{\rm H}\rightarrow \nu_\mu +n +n$ process,
calculated with the Bonn B potential~\cite{bonnb}
in the three-dimensional formalism of Ref.~\cite{edis3d}
and using the single nucleon current operator from Eqs.~(\ref{jnr01}) 
and~(\ref{jnrvec1}) 
for $F=\frac12$ (left panel)
and $F=\frac32$ (right panel) as a function
of the neutrino energy $E_\nu$.
The dashed curves show the plane wave results and the solid curves
are used for the full results.
Note that the average ``nucleon mass'' is used in the kinematics 
and in solving the Lippmann-Schwinger equations (see text for more details).
\label{fig05}}
\end{figure}

\begin{figure}
\includegraphics[width=7cm]{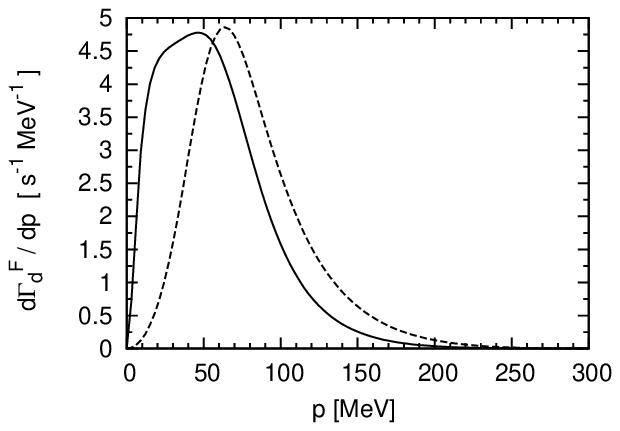}
\includegraphics[width=7cm]{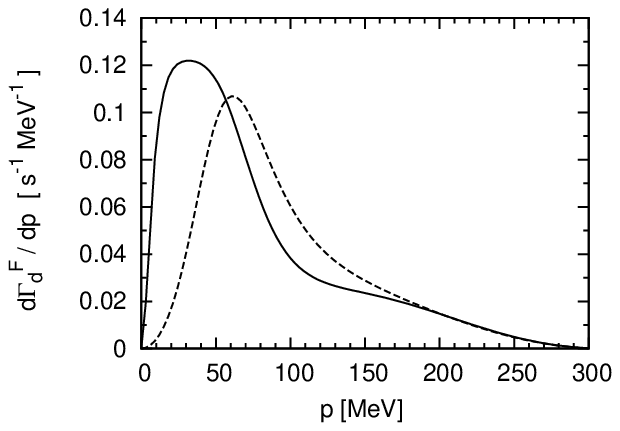}
\caption{The same as in Fig.~\ref{fig05} but 
given in the form of ${d\Gamma_d^F }/ {dp}$
and shown as a function
of the magnitude of the relative neutron-neutron momentum $p$.\label{fig1}}
\end{figure}

\begin{figure}
\includegraphics[width=7cm]{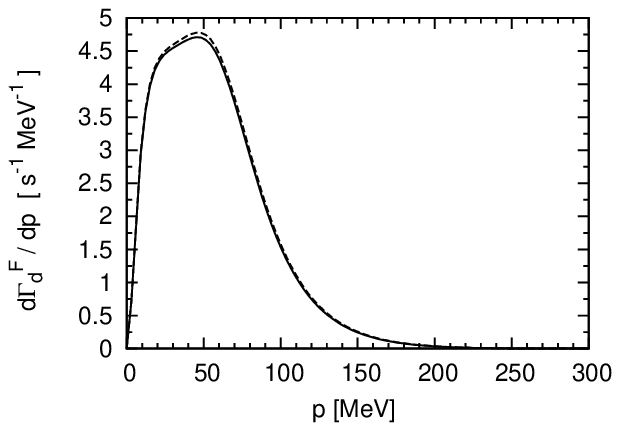}
\includegraphics[width=7cm]{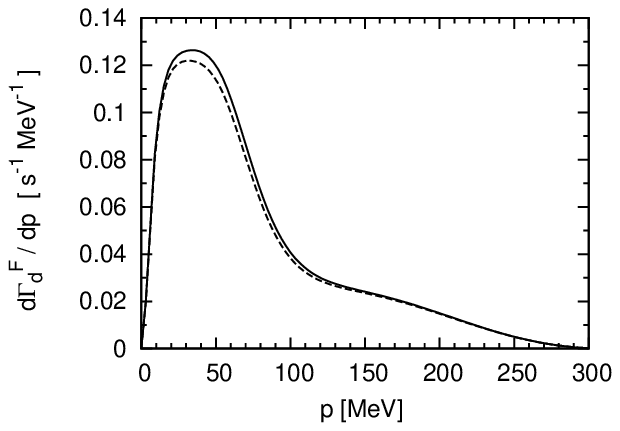}
\caption{Differential capture rate $ {d\Gamma_d^F }/ {dp} $
of the $\mu^-+^2{\rm H}\rightarrow \nu_\mu +n +n$ process calculated with the Bonn B potential~\cite{bonnb}
in the three-dimensional formalism of Ref.~\cite{edis3d}
for $F=\frac12$ (left panel)
and $F=\frac32$ (right panel) as a function
of the relative neutron-neutron momentum $p$.
The dashed (solid) curves show the full results obtained 
with the single nucleon current operator
without (with) the relativistic corrections.
Note that the average ``nucleon mass'' is used in the kinematics
and in solving the Lippmann-Schwinger equations (see text for more details).
\label{fig2}}
\end{figure}

From Eq.~(\ref{gnnf}) one can easily read out the differential capture rate 
$ d\Gamma_d^F /d E_\nu $. As shown in Fig.~\ref{fig05} this quantity 
soars in the vicinity of $E_\nu^{max,nn}$ (especially for the full results, 
which include the neutron-neutron final state interaction), 
which makes the observation
of dynamical effects quite difficult.
That is why the differential capture rate is usually shown 
as a function of the magnitude of the relative momentum.
The transition between $ d\Gamma_d^F /d E_\nu $
and $ d\Gamma_d^F /d p $ 
is given by Eq.~(\ref{enn}) and reads 
\begin{eqnarray}
\frac{ d\Gamma_d^F }{ d p } = 
\frac{ d\Gamma_d^F }{d E_\nu } \, \Big| \frac{dE_\nu} {dp } \Big| = 
\frac{ d\Gamma_d^F }{d E_\nu } \, \Big| \frac1{\frac{dp} {dE_\nu} } \Big| =
\frac { 4 p } { E_\nu + 2 M } \, \frac{ d\Gamma_d^F }{d E_\nu } \, .
\label{dgdf2}
\end{eqnarray}  
Our predictions shown in 
Figs.~\ref{fig05}, ~\ref{fig1} and~\ref{fig2}
are obtained in the three-dimensional formalism of Ref.~\cite{edis3d}, without any resort to partial wave
decomposition (PWD). These results for the Bonn B potential \cite{bonnb} can be used to 
additionally prove the convergence of other results based on partial waves.
These figures (and the corresponding numbers given in Table~\ref{tabnn})
show clearly that the doublet rate is dominant, as has
been observed before, for example in Ref.~\cite{prc83.014002}. Although the plane wave 
and full results for the total $F=\frac12$ and $F=\frac32$ rates are rather similar, 
the shapes of differential rates are quite different. The $1/M^2$ corrections 
in the current operator do not make significant contributions (see Fig.~\ref{fig2}) 
and the total rate is reduced only by about $ 2 $\% for $F=\frac12$
and raised by about $ 4 $\% for $F=\frac32$.

\begin{figure}
\includegraphics[width=7cm]{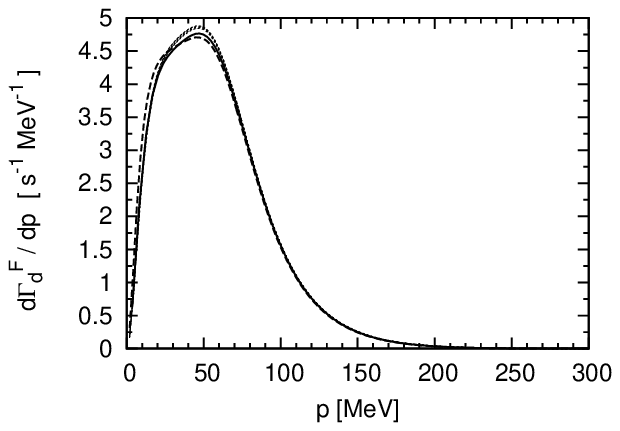}
\includegraphics[width=7cm]{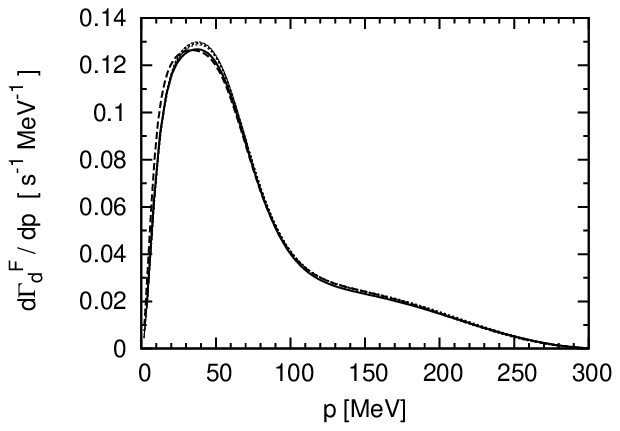}
\caption{Differential capture rate $ {d\Gamma_d^F }/ {dp} $
of the $\mu^-+^2{\rm H}\rightarrow \nu_\mu +n +n$ process 
calculated using standard PWD 
with various nucleon-nucleon potentials:
the AV18 potential \cite{av18} (solid curves),
the Bonn B potential \cite{bonnb} (dashed curves)
and the set of chiral NNLO potentials from Ref.~\cite{chiralnn} (bands)
for $F=\frac12$ (left panel)
and $F=\frac32$ (right panel) as a function
of the relative neutron-neutron momentum $p$.
Note that the bands are very narrow and thus appear practically as a curve.
All the partial wave states with $j \le 4$ have been included
in the calculations with the single nucleon current operator
containing the relativistic corrections.
Note that the average ``nucleon mass'' is used in the kinematics
and in solving the Lippmann-Schwinger equations (see text for more details).
\label{fig3}}
\end{figure}

In Fig.~\ref{fig3} we see that our predictions calculated with different nucleon-nucleon 
potentials lie very close to each other. We take the older Bonn B potential \cite{bonnb}, the 
AV18 potential \cite{av18} and five different parametrizations 
of the chiral next-to-next-to-leading order (NNLO) potential
from the Bochum-Bonn group \cite{chiralnn}. The corresponding total $F=\frac12$ rates vary 
only by about 
$2$\%, while the total $F=\frac32$ rates are even more stable.
It remains to be seen, if the same effects can be found with a more complicated
current operator.

The doublet and quadruplet total capture rates
are given in Table~\ref{tabnn} with the various nucleon-nucleon
potentials indicated above and the different approximations already
discussed for Figs.~\ref{fig05}-\ref{fig3}. 
The experimental data of
Refs.~\cite{Wan65,Ber73,Bar86,Car86} are
also shown. Since the experimental uncertainties for these data are
very large, no conclusion can be drawn from a comparison with them.
Note that within the similar framework developed in Ref.~\cite{prc83.014002},
by including the same single nucleon current operator mentioned above, 
we obtain $\Gamma_d^{F=1/2} =378$ s$^{-1}$ (235 s$^{-1}$ for the $^1S_0$ 
neutron-neutron partial wave), to be
compared with the value of 392 s$^{-1}$ of Table~\ref{tabnn}. The difference
of 14 s$^{-1}$ is due to 
(i) the use of the average ``nucleon mass'' in the Lippmann-Schwinger equation
for the $t$-matrix and final state kinematics    ($\approx$ 10 s$^{-1}$),
(ii) $j>2$ 2N partial wave contributions ($\approx$ 3 s$^{-1}$). 
Since for the pure neutron-neutron system we can use the true 
neutron mass, we have performed the corresponding momentum 
space calculation with $j\le2$ partial wave states and 
obtained $\Gamma_d^{F=1/2} =380$ s$^{-1}$
(237 s$^{-1}$ for the $^1S_0$ neutron-neutron partial wave), which proves 
a very good agreement with Ref.~\cite{prc83.014002}.

The above results have been calculated using PWD. In the case of the Bonn B
potential they have been compared with the predictions obtained employing the
three-dimensional scheme and an excellent agreement has been found.
The 2N momentum space partial wave states carry information 
about the magnitude of the relative momentum ($p$),
the relative angular momentum ($l$), spin ($s$) and total angular
momentum ($j$) with the corresponding projection ($m_j$).
This set of quantum numbers is supplemented by the 2N 
isospin ($t$) and its projection ($m_t$).
In order to avoid the cumbersome task of PWD of the many terms
in Eqs.~(\ref{1nr+rc01}) and (\ref{1nr+rcvec1}) 
we proceed in the same way as for the nuclear potentials 
in the so-called automatized PWD method  \cite{apwd1,apwd2}.
In the case of the single nucleon current operator
it leads to a general formula
\begin{eqnarray}  
&&\langle p (l s) j m_j \, t m_t \, {\bf P}_f \mid j_w (1) \mid \phi_d \, {\bf P}_i=0 \, m_d \, \rangle \ = \
\delta_{ t , 1 } \, 
\delta_{ m_t , -1 } \, 
\Big\langle  1 -1 \mid \tau_{-} (1) \mid 0  0 \, \Big\rangle 
\nonumber \\
&&c \left( l , s , j ;  m_l,  m_j - m_l, m_j \,  \right) \,
\sum\limits_{l_d = 0 , 2}  \,
\sum\limits_{m_{l_d}}  
c \left( l_d , 1 , 1 ;  m_{l_d},  m_d - m_{l_d} , m_d \,  \right) \,
\nonumber \\
&&\sum\limits_{m_1}  
c \left( \frac12 , \frac12 , s  ;  m_1,  m_j - m_l - m_1  , m_j - m_l  \,  \right) \,
\nonumber \\
&&\sum\limits_{m_{1_b}}  
c \left( \frac12 , \frac12 , 1  ;  m_{1_b},  m_d - m_{l_d} - m_{1_d}   , m_d - m_{l_d}   \,  \right) \,
\nonumber \\
&&\delta_{ m_j - m_l - m_1  , m_d - m_{l_d} - m_{1_d} \, } \,
\nonumber \\
&&\int d \hat{\bf p} \,
Y^*_{l \, m_l } \left( \hat{\bf p} \right) \,
Y_{l_d \, m_{l_d} } \left( \widehat{{\bf p} - \frac12 {\bf Q} \, } \right) \,
\varphi_{l_d} \left( \mid {\bf p } - \frac12 {\bf Q} \mid \, \right) \,
\nonumber \\
&&\Big\langle  \frac12 m_1 \mid \Big\langle \, {\bf p} + \frac12{\bf P}_f  
\mid j_w^{\rm spin} (1) \mid  {\bf p} - \frac12{\bf P}_f + {\bf P}_i \, \big\rangle 
\mid \frac12  m_{1_d}  \,  \Big\rangle \,
\label{pwdj1d}
\end{eqnarray}  
where $ {\bf Q} \equiv {\bf P}_f - {\bf P}_i $
and the deuteron state contains two components
\begin{eqnarray}  
\mid \phi_d \, m_d \, \rangle  =
\sum\limits_{l_d= 0, 2} \,
\int dp p^2  \, \mid p ( l_d 1 ) 1 m_d \, \rangle \mid  0 0  \, \rangle \, \varphi_{l_d} \left( p \right) \, .
\label{pwdj1d2}
\end{eqnarray}  
Using software for symbolic algebra, 
for example {\em Mathematica}$^\registered$~\cite{math}, 
we easily 
prepare momentum dependent spin matrix elements 
\begin{eqnarray}  
\Big\langle  \frac12 m^{\, \prime}  \mid \Big\langle {\bf p}_1^{\ \prime} 
\mid j_w^{\rm spin} (1) \mid  
{\bf p}_1 \, \Big\rangle \mid \frac12 m \, \Big\rangle \,
\label{pwdj1d3}
\end{eqnarray}  
for any type of the single nucleon operator. 
The calculations have been performed including
all partial wave states with 
$ j \le 4$. We typically use 40 $E_\nu$ points 
and 50 $\theta_p$ values to achieve fully converged results.
Note that in Ref.~\cite{prc83.014002}, a standard multipole expansion was 
obtained retaining all $j\leq 2$ and $l\leq 3$ neutron-neutron partial waves, and the 
integration over $p$ ($\theta_p$) was performed with 30 ($\sim$ 10) 
integration points.

\begin{table}
\caption{Doublet ($F=1/2$) and quadruplet ($F=3/2$) capture rates 
for the $\mu^- + ^2{\rm H} \rightarrow \nu_\mu + n + n $ reaction
calculated with various nucleon-nucleon potentials and the single nucleon
current operator without and with the relativistic corrections (RC).
Plane wave results (PW) and results obtained with the rescattering
term in the nuclear matrix
element (full) are shown. 
Note that the average ``nucleon mass'' is used in the kinematics
and in solving the Lippmann-Schwinger equations (see text for more details).
The available experimental
data are from Refs.~\cite{Wan65,Ber73,Bar86,Car86}.\label{tabnn}}
\begin{tabular}{l|c|c|c|c}
\hline\hline
& \multicolumn{4}{c}{Capture rate $\Gamma_d^F$ in s$^{-1}$} \\ \cline{2-5}
& \multicolumn{2}{c|}{$F=1/2$} & \multicolumn{2}{|c}{$F=3/2$} \\ \cline{2-5}
\hline
nucleon-nucleon force and dynamics & PW & full & PW & full \\ 
\hline
Bonn B, without RC  &  369 & 403  & 10.0 & 11.7 \\
Bonn B, with RC     &  363 & 396  & 10.4 & 12.2 \\
AV18, with RC       &  361 & 392  & 10.2 & 12.0 \\
chiral NNLO potential version 1 with RC  &  367 & 399 & 10.5 & 12.2 \\
chiral NNLO potential version 2 with RC  &  364 & 394 & 10.4 & 12.2 \\
chiral NNLO potential version 3 with RC  &  365 & 397 & 10.5 & 12.2 \\
chiral NNLO potential version 4 with RC  &  367 & 399 & 10.4 & 12.2 \\
chiral NNLO potential version 5 with RC  &  364 & 396 & 10.4 & 12.2 \\
\hline
experimental results: & \multicolumn{2}{|r|}{} & \multicolumn{2}{r}{}\\
I.-T. Wang {\it et al.}   \cite{Wan65}     &  \multicolumn{2}{c|}{365 $\pm$ 96} 
& \multicolumn{2}{|c} {} \\
A. Bertin {\it et al.} \cite{Ber73}          &  \multicolumn{2}{c|}{445 $\pm$ 60} & \multicolumn{2}{|c} {} \\
G. Bardin {\it et al.} \cite{Bar86}          &  \multicolumn{2}{c|}{470 $\pm$ 29} & \multicolumn{2}{|c} {} \\
M. Cargnelli {\it et al.} \cite{Car86}       &  \multicolumn{2}{c|}{409 $\pm$ 40} & \multicolumn{2}{|c} {} \\ 
\hline\hline
\end{tabular}
\end{table}

\section{Results for the $\mu^- + ^3{\rm He} \rightarrow \nu_\mu + ^3{\rm H} $ reaction}
\label{section4}

In this case we deal with simple two-body kinematics
and we can compare the neutrino energy calculated nonrelativistically 
and using relativistic equations. 
The relativistic result, based on 
\begin{eqnarray}
M_\mu + M_{^3{\rm He}} = E_\nu + \sqrt{ E_\nu^2 + M_{^3{\rm H}}^2 \, }
\label{3h1rel}
\end{eqnarray}
reads
\begin{eqnarray}
\left( E_\nu \right)^{rel} = \frac{ \left( M_{^3{\rm He}} + M_\mu \, \right)^2 - M_{^3{\rm H}}^2   } 
{ 2 \, \left( M_{^3{\rm He}} + M_\mu \, \right)   } \, .
\label{3h2rel}
\end{eqnarray}
In the nonrelativistic case, we start with 
\begin{eqnarray}
M_\mu + M_{^3{\rm He}} = E_\nu + M_{^3{\rm H}} + \frac{ E_\nu^2 } { 2 M_{^3{\rm H}}  }
\label{3h1nrl}
\end{eqnarray}
and arrive at
\begin{eqnarray}
\left( E_\nu \right)^{nrl} = - M_{^3{\rm H}} 
+ \sqrt{ M_{^3{\rm H}} \left( -M_{^3{\rm H}} + 2 \left( M_{^3{\rm He}} + M_\mu \, \right)  \right) }
 \, .
\label{3h2nrl}
\end{eqnarray}
Again the obtained numerical values,
$ \left( E_\nu \right)^{rel}$  = 103.231 MeV
and
$ \left( E_\nu \right)^{nrl}$  = 103.230 MeV,
are very close to each other.

For this case we do not consider the ($F=0$ and $F=1$) hyperfine states 
in $ ^3{\rm He}$ and calculate 
directly 
\begin{eqnarray}
&&\Gamma_{^3{\rm H}} = \frac12 G^2 \frac1{( 2 \pi )^2 } \, {\cal R} \,
\frac { \left(  2 M^\prime_{^3{\rm He}} \alpha \, \right)^3 } {\pi  } \, \rho \, 
\nonumber \\
&&4 \pi \,
\frac12 \sum\limits_{m_{^3{\rm He}}} \sum\limits_{m_{^3{\rm H}} } 
\Big( 
\left| N^0 (m_{^3{\rm H}}, m_{^3{\rm He}}  \, ) \, \right |^2 \, + \, 
\left| N_z (m_{^3{\rm H}}, m_{^3{\rm He}}  \, ) \, \right |^2 \, + \, 
\nonumber \\
&&2 \left| N_{-1} (m_{^3{\rm H}}, m_{^3{\rm He}} \, ) \, \right |^2 \, + \, 
2 {\rm Re} \left( N^0 (m_{^3{\rm H}}, m_{^3{\rm He}} \, )  
\left( N_z (m_{^3{\rm H}}, m_{^3{\rm He}} \, ) \right)^*   \right ) \, \Big) \,  ,
\label{g3h}
\end{eqnarray}  
where 
the factor $ \frac { \left( 2  M^\prime_{^3{\rm He}} \alpha \, \right)^3 } {\pi  } $, like 
in the deuteron case, comes from the 
$K$-shell atomic wave function and 
$ M^\prime_{^3{\rm He}}  = \frac { M_{^3{\rm He}} M_\mu } { M_{^3{\rm He}} + M_\mu }$. 
Also in this case one can fix the direction of the neutrino momentum
(our choice is $ {\hat {\bf p}}_\nu = - {\hat {\bf z}} $) 
and the angular integration yields just $ 4 \pi$.
The phase space factor $ \rho$ is 
\begin{eqnarray}
\rho = \frac {E_\nu^2 } {1 + \frac {E_\nu}{\sqrt{ E_\nu^2 + M_{^3{\rm H}}^2 \, } }  } \, 
\approx \,
E_\nu^2 \,  \left( 1 - \frac {E_\nu} {M_{^3{\rm H}} } \,   \right)  \, .
\label{rho3h}
\end{eqnarray}  
The additional factor ${\cal R}$ accounts for the finite 
volume of the $^3$He charge and we 
assume that ${\cal R}= 0.98 $ \cite{prc83.014002}.  
(The corresponding factor in the deuteron case has been found to be very 
close to $1$ \cite{prc83.014002} and thus is omitted.)
Now, of course, the nuclear matrix elements involve the initial
$^3$He and final $^3$H states:
\begin{eqnarray}  
N^\lambda (m_{^3{\rm H}}, m_{^3{\rm He}}  \, )  \, \equiv \, 
\BA{\Psi_{^3{\rm H}} \, {\bf P}_f=-{\bf p}_\nu  \, m_{^3{\rm H}} \, } \, 
j_w^\lambda
\, \KT{\Psi_{^3{\rm He}} \, {\bf P}_i=0 \, m_{^3{\rm He}} \, } 
\label{n3h}
\end{eqnarray}  
and many-nucleon contributions are expected  in $j_w^\lambda$ 
as given in Eq.~(\ref{jwmany}). 

Our results for this process are given in Table~\ref{tab3h}. 
They are based on various 3N Hamiltonians
and the single nucleon current operator. 
Only in the last line we show a result, where on top of the single
nucleon contributions 2N operators are added to the current
operator $ j_w^\lambda $. We use the meson-exchange currents from 
Ref.~\cite{prc63.015801} (Eqs.~(4.16)--(4.39), without $\Delta$-isobar 
contributions). Among the 2N operators listed in that reference, there 
are so-called non-local structures (like the one in Eq.~(4.37)) 
and their numerical 
implementation in our 3N calculations is quite involved. The local
structures can be treated easily as described for example 
in Refs.~\cite{physrep,rozpedzik}. 
Our two last results from Table~\ref{tab3h} 
(1324 s$^{-1}$ and  1386 s$^{-1}$),
should be compared with the PS (1316 s$^{-1}$) and Mesonic (1385 s$^{-1}$) predictions 
from Table~X of Ref.~\cite{prc83.014002}, although not all the details 
of the calculations are the same. 
The experimental value 
for this capture rate is known with a rather good accuracy 
($\Gamma_{exp} =  (1496 \pm 4)$  s$^{-1}$~\cite{Acker98})
so one can expect that the effects of 2N operators 
exceed $11$\%.
At least for this process, they are more important than the 3N force effects.
The latter ones amount roughly to $2$\% only.
This dependence on the 3N interaction was already 
observed in Ref.~\cite{Mar02}, where it was shown that the total capture rate 
scales approximately linearly with the trinucleon binding energy.

In the 3N case we employ PWD
and use our standard 3N basis $\mid  p q \bar{\alpha }\, J m_J ; T m_T \, \rangle $
\cite{physrep}, where $p$ and $q$ are magnitudes of the relative Jacobi
momenta and $ \bar{\alpha}$ is a set of discrete quantum numbers.
Note that the $\mid  p q \bar{\alpha } \, J m_J ; T m_T \, \rangle $ states are 
already antisymmetrized in the $(2,3)$ subsystem.
Also in this case we have derived a general formula for PWD
of the single nucleon current operator:
\begin{eqnarray}  
&&\langle p q \bar{\alpha} J m_J ; T m_T \, {\bf P}_f \mid j_w (1) \mid \Psi_{^3{\rm He}} \, {\bf P}_i=0 \, m_{^3{\rm He}} \, \rangle = 
\nonumber \\
&&\sum\limits_{\bar{\alpha}_b} 
\delta_{ l  , l_b} \, 
\delta_{ s , s_b} \, 
\delta_{ j , j_b} \, 
\delta_{ t , t_b} \, 
\delta_{ m_T , -\frac12 } \, 
\Big\langle  \left( t \frac12 \right) T \, -\frac12 \mid \tau_{-} (1) \mid 
 \left( t_b \frac12 \right) \frac12 \, \frac12 \, \Big\rangle 
\nonumber \\
&&\sum\limits_{m_j}  
c \left( j,  I, J ; m_j , m_J - m_j , m_J \,  \right) \,
c \left( j_b,  I_b, \frac12 ; m_j , m_{^3{\rm He}} - m_j , m_{^3{\rm He}} \,  \right) \,
\nonumber \\
&&\sum\limits_{m_\lambda}  
c \left( \lambda , \frac12 , I ; m_\lambda , m_J - m_j - m_\lambda , m_J - m_j \,  \right) \,
\nonumber \\
&&\sum\limits_{m_{\lambda_b}}  
c \left( \lambda_b , \frac12 , I_b ; m_{\lambda_b} , m_{^3{\rm He}} - m_{j_b}  - m_{\lambda_b} , m_{^3{\rm He}} - m_{j_b} \,  \right) \,
\nonumber \\
&&\int d \hat{\bf q} \,
Y^*_{\lambda\, m_\lambda } \left( \hat{\bf q} \right) \,
Y_{\lambda_b \, m_{\lambda_b} } \left( \widehat{{\bf q} - \frac23 {\bf Q} \, } \right) \,
\phi_{\bar{\alpha}_b} \left( p , \mid {\bf q} - \frac23 {\bf Q} \mid \, \right)
\nonumber \\
&&\Big\langle  \frac12 m_J - m_j - m_\lambda \mid \Big\langle \, {\bf q} + \frac13{\bf P}_f  
\mid j_w^{\rm spin} (1) \mid  {\bf q} - \frac23{\bf P}_f + {\bf P}_i \, \big\rangle 
\mid \frac12 m_{^3{\rm He}} - m_{j_b}  - m_{\lambda_b} \,  \Big\rangle \,
\label{pwdj1}
\end{eqnarray}  
where, as in the 2N space, $ {\bf Q} \equiv {\bf P}_f - {\bf P}_i $.
We encounter again the essential spin matrix element
\begin{eqnarray}  
\Big\langle  \frac12 m^{\, \prime} \Big|  \Big\langle \, {\bf p}_1^{\ \prime}  
\Big| j_w^{\rm spin} (1) \Big| {\bf p}_1 \, \big\rangle \,
\Big| \frac12 m \Big\rangle \, 
\label{pwdj12}
\end{eqnarray}
of the single nucleon current operator, which
is calculated using software for symbolic algebra.
The initial 3N bound state is given as
\begin{eqnarray}  
\mid \Psi_{^3{\rm He}} \, m_{^3{\rm He}} \, \rangle  =
\sum\limits_{\bar{\alpha}_b} \,
\int dp p^2 \int dq q^2 \, 
\Big| p q \bar{\alpha}_b \, \frac12 \, m_{^3{\rm He}} \, ;
\frac12 \frac12 \Big\rangle \, 
\phi_{\bar{\alpha}_b} \left( p , q \right) \, .
\label{pwdj13}
\end{eqnarray}  
In our calculations we have used 34 (20) points for integration over $p$
($q$), and 
34 partial wave states corresponding to $j \le 4$.

\begin{table}
\caption{Total capture rate $\Gamma$ 
for the 
$\mu^- + ^3{\rm He} \rightarrow \nu_\mu + ^3{\rm H}$ reaction
calculated with the single nucleon
current operator and various nucleon-nucleon potentials. In the last
two lines the rates are obtained 
employing the AV18~\cite{av18} nucleon-nucleon 
and the Urbana~IX 3N potential~\cite{urbana}, and adding,
in the last line, some selected 2Ns current operators to the
single nucleon current (see text for more explanations).
\label{tab3h}}
\begin{ruledtabular}
\begin{tabular}{lc}
Three-nucleon Hamiltonian & Capture rate $\Gamma$ in s$^{-1}$ \\
\hline
Bonn B  &  1360  \\
chiral NNLO version 1 & 1379 \\
chiral NNLO version 2 & 1312 \\
chiral NNLO version 3 & 1350 \\
chiral NNLO version 4 & 1394 \\
chiral NNLO version 5 & 1332 \\
AV18  &  1353  \\
\hline
AV18 + Urbana~IX  &  1324  \\
AV18 + Urbana~IX with MEC \cite{prc63.015801}  &  1386 
\end{tabular}
\end{ruledtabular}
\end{table}

\section{Results for the $\mu^- + ^3{\rm He} \rightarrow \nu_\mu + n + d $ 
and $\mu^- + ^3{\rm He} \rightarrow \nu_\mu + n + n + p $ reactions}
\label{section5}

The kinematics of the 
$\mu^-+ ^3{\rm He} \rightarrow \nu_\mu + n + d $ 
and
$\mu^-+ ^3{\rm He} \rightarrow \nu_\mu + n + n + p $ 
reactions is formulated in the same way
as for the $\mu^-+ ^2{\rm H} \rightarrow \nu_\mu + n + n $
process in Sec.~\ref{section3}.
The maximal neutrino energies 
for the two-body and three-body captures of the muon atom 
are evaluated as

\begin{eqnarray}
\left( E_\nu^{max,nd} \, \right)^{rel} &=&
\frac{({M_{^3{\rm He}}}-{M_d}+{M_\mu}-{M_n})
   ({M_{^3{\rm He}}}+{M_d}+{M_\mu}+
   {M_n})}{2 ({M_{^3{\rm He}}}+{M_\mu})} \, ,
\label{Enumaxrelnd} \\
\left( E_\nu^{max,nnp} \, \right)^{rel} &=&
\frac{{M_{^3{\rm He}}}^2+2 {M_{^3{\rm He}}}
   {M_\mu}+{M_\mu}^2-(2
   {M_n}+{M_p})^2}{2
   ({M_{^3{\rm He}}}+{M_\mu})} \, ,
\label{Enumaxrelnnp} \\
\left( E_\nu^{max,nd} \, \right)^{nrl} &=&
\sqrt{
( M_d + M_n ) ( 2 M_{^3{\rm He}} + 2 M_\mu - M_d - M_n ) 
}-{M_d}-{M_n} \, ,
\label{Enumaxnrlnd} \\
\left( E_\nu^{max,nnp} \, \right)^{nrl} &=&
\sqrt{
( M_p + 2 M_n ) ( 2 M_{^3{\rm He}} + 2 M_\mu - 2 M_n - M_p ) 
}-2 {M_n}-{M_p} \, .
\label{Enumaxnrlnnp}
\end{eqnarray}
The numerical values are the following:
$\left( E_\nu^{max,nd} \, \right)^{rel} $ = 97.1947 MeV,
$\left( E_\nu^{max,nd} \, \right)^{nrl} $ = 97.1942 MeV,
$\left( E_\nu^{max,nnp} \, \right)^{rel} $ = 95.0443 MeV
and
$\left( E_\nu^{max,nnp} \, \right)^{nrl} $ = 95.0439 MeV.

\begin{figure}
\includegraphics[width=8cm]{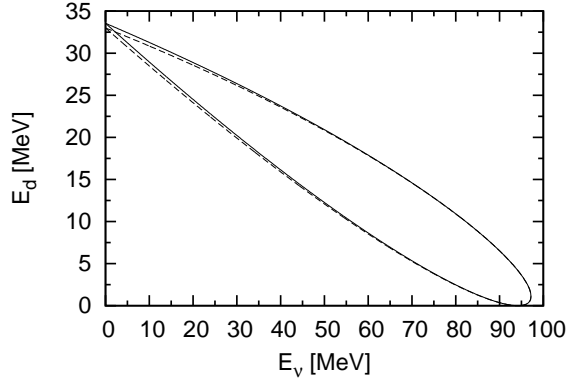}
\caption{The kinematically allowed region in the 
$E_\nu - E_d$ plane calculated relativistically 
(solid curve) and
nonrelativistically (dashed curve) for the
$\mu^-+ ^3{\rm He} \rightarrow \nu_\mu + n + d $ process. \label{enu_ed}}
\end{figure}

\begin{figure}
\includegraphics[width=8cm]{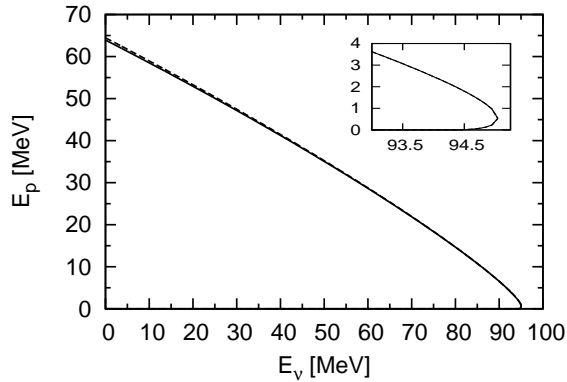}
\caption{The kinematically allowed region in the 
$E_\nu - E_p$ plane calculated relativistically 
(solid curve) and
nonrelativistically (dashed curve) for the
$\mu^-+ ^3{\rm He} \rightarrow \nu_\mu + n + d $ process. \label{enu_ep}}
\end{figure}

The kinematically allowed region 
in the $E_\nu - E_d$ plane for the 
two-body break-up of $^3$He is shown in Fig.~\ref{enu_ed}.
We show the curves based on the relativistic and nonrelativistic kinematics.
They essentially overlap except for the very small neutrino energies.
The same is also true for the three-body break-up as demonstrated
in Fig.~\ref{enu_ep}. Up to a certain $E_\nu$ value, which we denote by
$E_\nu^{2sol}$, the minimal proton kinetic energy is zero. 
The minimal proton kinetic energy is greater than zero
for $ E_\nu > E_\nu^{2sol}$. Even this very detailed shape 
of the kinematical domain can be calculated 
nonrelativistically with high accuracy (see also the inset in Fig.~\ref{enu_ep}).
The  values of $E_\nu^{2sol}$
based on the relativistic kinematics, 
\begin{equation}
\left( E_\nu^{2sol} \, \right)^{rel} =
\frac{
( M_{^3{\rm He}} + M_\mu ) ( M_{^3{\rm He}} + M_\mu - 2 M_p ) - 4 {M_n}^2+{M_p}^2
}{2 ({M_{^3{\rm He}}}+{M_\mu}-{M_p})}
\label{Enu2solrel}
\end{equation}
and nonrelativistic kinematics, 
\begin{equation}
\left( E_\nu^{2sol} \, \right)^{nrl} =
2 \left(\sqrt{{M_{^3{\rm He}}}
   {M_n}+{M_\mu}
   {M_n}-{M_n}^2-
   {M_n}
   {M_p}}-{M_n}\right) \, ,
\label{Enu2solnrl}
\end{equation}
yield very similar numerical values, 
94.2832 MeV
and
94.2818 MeV, respectively.

In Ref.~\cite{romek} we performed the first calculations for 
the $\mu^- + ^3{\rm He} \rightarrow \nu_\mu + n + d $ reaction
taking into account only nucleon-nucleon forces but including 
final state interactions.
We analyzed some experimental data \cite{datainromek1,datainromek2}
and found large effects of final state interactions. 
In the present paper 
we calculate the total capture rate for 
the two-body and three-body break-up reactions 
and analyze more complete 
data sets from Refs.~\cite{datainromek1,datainromek2} 
and Ref.~\cite{pra69.012712}. The two-body and three-body nuclear
scattering states are here obtained including a 3N force. 
To this end we use the experience from our studies on electromagnetic reactions
(see for example Refs.~\cite{physrep,romek2}). 

The crucial matrix elements
\begin{eqnarray}  
N_{nd}^\lambda (m_n, m_d , m_{^3{\rm He}}  \, )  \, \equiv \, 
\BA{\Psi_{nd}^{(-)}  \, {\bf P}_f=-{\bf p}_\nu  \, m_n \, m_d } \, 
j_w^\lambda
\, \KT{\Psi_{^3{\rm He}} \, {\bf P}_i=0 \, m_{^3{\rm He}} \, } 
\label{nnd}
\end{eqnarray}  
and
\begin{eqnarray}  
N_{nnp}^\lambda (m_1, m_2 , m_p , m_{^3{\rm He}}  \, )  \, \equiv \, 
\BA{\Psi_{nnp}^{(-)}  \, {\bf P}_f=-{\bf p}_\nu  \, m_1 \, m_2 \, m_p} \, 
j_w^\lambda
\, \KT{\Psi_{^3{\rm He}} \, {\bf P}_i=0 \, m_{^3{\rm He}} \, } 
\label{nnnp}
\end{eqnarray}  
are calculated in two steps.
First we solve a Faddeev-like equation 
for the auxiliary state $ \KT{ U^\lambda \, } $ for each considered 
neutrino energy:
\begin{eqnarray} 
\KT{ U^\lambda \, } = \Big[ t G_0 \, + \, \frac12 ( 1 + P ) V_4^{(1)}  G_0  ( 1 + t G_0 \, ) \, \Big] ( 1 + P ) j_w^\lambda \KT{\Psi_{^3{\rm He}} \, } 
\nonumber \\  
+ \  \Big[ t G_0 P \, + \, \frac12 ( 1 + P ) V_4^{(1)}  G_0  ( 1 + t G_0 P \, ) \, \Big] \KT{ U^\lambda \, }  \, ,
\label{u}
\end{eqnarray}  
where $ V_4^{(1)} $ is a part of the 3N force symmetrical under the exchange of nucleon 2 and ~3,
$G_0$ is the free 3N propagator and $t$ is the 2N $t$-operator
acting in the $(2,3)$ subspace. Further $P$ is the permutation operator built from the 
transpositions $P_{ij}$ exchanging nucleons $i$ and $j$:  
\begin{eqnarray} 
P = P_{12} P_{23} + P_{13} P_{23} \, .
\label{p}
\end{eqnarray}  
In the second step the nuclear matrix elements are calculated 
by simple quadratures:
\begin{eqnarray}  
N_{nd}^\lambda (m_n, m_d , m_{^3{\rm He}}  \, )  &=& 
\BA{\phi_{nd} \, {\bf q}_0 \, m_n \, m_d } \, ( 1 + P ) j_w^\lambda \KT{\Psi_{^3{\rm He}} \, }
\nonumber \\
&+&
\BA{\phi_{nd} \, {\bf q}_0 \, m_n \, m_d } \,  P  \KT{ U^\lambda \, }  \, ,
\label{nnd2} \\
N_{nnp}^\lambda (m_1, m_2 , m_p , m_{^3{\rm He}}  \, )  &=&
\BA{\phi_{nnp} \, {\bf p} \, {\bf q} \, m_1 \, m_2 \, m_p \, } \, ( 1 + P ) j_w^\lambda \KT{\Psi_{^3{\rm He}} \, }
\nonumber \\
&+&
\BA{\phi_{nnp} \, {\bf p} \, {\bf q} \, m_1 \, m_2 \, m_p \, } \, t G_0  ( 1 + P ) j_w^\lambda \KT{\Psi_{^3{\rm He}} \, }
\nonumber \\
&+&
\BA{\phi_{nnp} \, {\bf p} \, {\bf q} \, m_1 \, m_2 \, m_p \, } \,  P \KT{ U^\lambda \, }
\nonumber \\
&+&
\BA{\phi_{nnp} \, {\bf p} \, {\bf q} \, m_1 \, m_2 \, m_p \, } \,  t G_0 P \KT{ U^\lambda \, }  \, .
\label{nnnp2}
\end{eqnarray}  
Here $\KT{\phi_{nd} \, {\bf q}_0 \, m_n \, m_d }$
is a product state of the deuteron wave function and a momentum eigenstate
of the spectator nucleon characterized by the relative momentum vector $ {\bf q}_0$, while
$\KT{\phi_{nnp} \, {\bf p} \, {\bf q} \, m_1 \, m_2 \, m_p \, }$
is a product state of two free motions in the 3N system 
given by Jacobi relative momenta $ {\bf p} $ and ${\bf q} $,
antisymmetrized in the $(2,3)$ subsystem.
Equations (\ref{u}), (\ref{nnd2}) and (\ref{nnnp2}) simplify significantly,
when $ V_4^{(1)} = 0$ \cite{romek2}. 

Finally we give our formulas for the total capture rates.
Like for the $\mu^- + ^3{\rm He} \rightarrow \nu_\mu + ^3{\rm H}$ reaction,
also for the two break-up channels these quantities
are calculated directly and the hyperfine states in $^3{\rm He}$ are not considered.
In the case of the two-body break-up it reads:
\begin{eqnarray}
&&\Gamma_{nd} = \frac12 G^2 \frac1{( 2 \pi )^2 } \, 
{\cal R} \, \frac { \left(  2 M^\prime_{^3{\rm He}} \alpha \, \right)^3 } {\pi  } \, 4 \pi \, 
\nonumber \\
&&\int\limits_0^{E_\nu^{max,nd}} \, dE_\nu E_\nu^2  \,
\frac23 M q_0  \, \frac13 \, 
\int\limits_{0}^{\pi} d \theta_{q_0} \sin \theta_{q_0} \, 2 \pi \, 
\nonumber \\
&&\frac12 \sum\limits_{m_{^3{\rm He}}} \sum\limits_{m_{n} , m_d } 
\Big( 
\left| N_{nd}^0 (m_{n}, m_d, m_{^3{\rm He}}  \, ) \, \right |^2 \, + \, 
\left| N_{nd, \, z}  (m_{n}, m_d, m_{^3{\rm He}}  \, ) \, \right |^2 \, + \, 
\nonumber \\
&&2 \left| N_{nd, \, -1} (m_{n}, m_d, m_{^3{\rm He}} \, ) \, \right |^2 \, + \, 
\nonumber \\
&&2 {\rm Re} \left( N_{nd}^0 (m_{n}, m_d, m_{^3{\rm He}} \, )  
\left( N_{nd, \, z } (m_{n}, m_d, m_{^3{\rm He}} \, ) \right)^*   \right ) \, \Big) \,  ,
\label{gnd}
\end{eqnarray}  
where we used the same arguments as before to simplify the angular integrations. The
energy conservation is expressed in terms of the relative neutron-deuteron momentum 
\begin{eqnarray}
{\bf q}_0  \equiv \frac23 \left(  {\bf p}_n - \frac12 {\bf p}_d \,  \right)  \, ,
\label{q0}
\end{eqnarray} 
yielding 
\begin{eqnarray}
M_\mu + M_{^3{\rm He}} \approx E_\nu + M_n + M_d 
+ \frac34 \frac{ {\bf q}_0^{\ 2}} {M} + \frac16 \frac{E_\nu^2 } {M} \, ,
\label{eq0}
\end{eqnarray}  
where we neglect the deuteron binding energy.
For the $ \mu^- + ^3{\rm He} \rightarrow \nu_\mu + n + n + p $ reaction 
we obtain in a similar way:
\begin{eqnarray}
&&\Gamma_{nnp} = \frac12 G^2 \frac1{( 2 \pi )^2 } \, 
{\cal R} \, \frac { \left(  2 M^\prime_{^3{\rm He}} \alpha \, \right)^3 } {\pi  } \, 4 \pi \, 
\nonumber \\
&&\int\limits_0^{E_\nu^{max,nnp}} \, dE_\nu E_\nu^2  \,
\frac23 M q  \, \frac13 \, 
\int\limits_{0}^{\pi} d \theta_{q} \sin \theta_{q} \, 2 \pi \, 
\int\limits_{0}^{\pi} d \theta_{p} \sin \theta_{p} \, \int\limits_{0}^{2 \pi} d \phi_{p} \, 
\int\limits_0^{p^{max}} \, dp p^2  \,
\nonumber \\
&&\frac12 \sum\limits_{m_{^3{\rm He}}} \sum\limits_{m_1, m_2 , m_p } 
\Big( 
\left| N_{nnp}^0 (m_1, m_2, m_p, m_{^3{\rm He}}  \, ) \, \right |^2 \, + \, 
\left| N_{nnp, \, z}  (m_1, m_2, m_p, m_{^3{\rm He}}  \, ) \, \right |^2 \, + \, 
\nonumber \\
&&2 \left| N_{nnp, \, -1} (m_1, m_2, m_p, m_{^3{\rm He}} \, ) \, \right |^2 \, + \, 
\nonumber \\
&&2 {\rm Re} \left( N_{nnp}^0 (m_1, m_2, m_p, m_{^3{\rm He}} \, )  
\left( N_{nnp, \, z } (m_1, m_2, m_p, m_{^3{\rm He}} \, ) \right)^*   \right ) \, \Big) \, .
\label{gnnp}
\end{eqnarray}  
The energy conservation is expressed in terms of the Jacobi relative 
momenta ${\bf p}$ and ${\bf q}$ 
\begin{eqnarray}
{\bf p}  \equiv \frac12 \left(  {\bf p}_1 - {\bf p}_2 \,  \right) \, , \nonumber \\
{\bf q}  \equiv \frac23 \left(  {\bf p}_p - \frac12 \left(  {\bf p}_1 +  {\bf p}_2 \,  \right) \,  \right) \, ,
\label{pq}
\end{eqnarray} 
which leads to 
\begin{eqnarray}
M_\mu + M_{^3{\rm He}} \approx E_\nu + 3 M
+ \frac{ {\bf p}^{\ 2}} {M} 
+ \frac34 \frac{ {\bf q}^{\ 2}} {M} 
+ \frac16 \frac{E_\nu^2 } {M} \, .
\label{epq}
\end{eqnarray}  

\begin{figure}
\includegraphics[width=7cm]{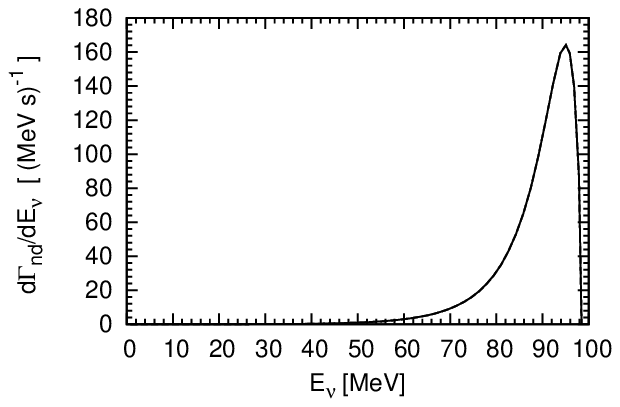}
\includegraphics[width=7cm]{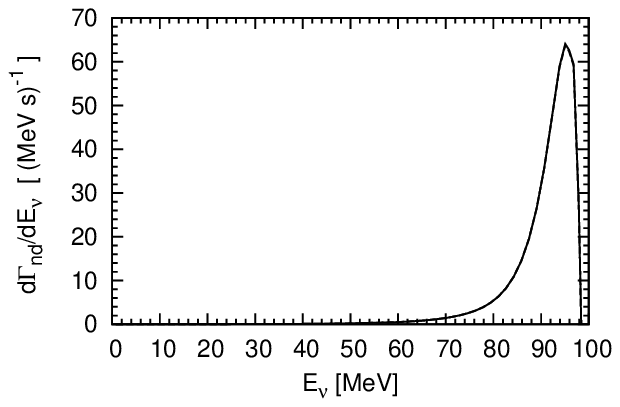}
\caption{The differential capture rates $ {d\Gamma }_{nd}/ {dE_\nu} $
for the $ \mu^- + ^{3}{\rm He} \rightarrow \nu_\mu + n + d$ 
process calculated with the AV18 potential \cite{av18} and the
single nucleon current operator
as a function of the muon neutrino energy, using the symmetrized plane wave 
(left panel) and a full solution of Eq.~(\ref{u})
with $ V_4^{(1)} =0 $
(right panel). The curves representing results of the calculations
employing all partial wave states with $j \le 3$ ($j \le 4$) in the 2N subsystem 
are depicted with dashed (solid) curves.
The maximal total 3N angular momentum is $J_{max} = \frac92$. \label{fig4}}
\end{figure}

We start the discussion of our predictions with Fig.~\ref{fig4}, 
where for the 
$ \mu^- + ^3{\rm He} \rightarrow \nu_\mu + n + d$ reaction
we compare results of calculations employing all partial wave 
states with the total subsystem angular momentum $j \le 3$ and $j \le 4$. 
Both the (symmetrized) plane wave and full results show a very good convergence
and in practice it is sufficient to perform calculations with $j \le 3$.
We refer the reader to Ref.~\cite{physrep} for the detailed definitions of 
various 3N dynamics.
The convergence with respect to the total 3N angular momentum $J$
will be discussed in Sec.~\ref{section6}. The differential capture rates 
$d \Gamma_{nd}/dE_{\nu_\mu}$ rise very slowly with the neutrino energy and
show a strong maximum in the vicinity
of the maximal neutrino energy. (At the very maximal neutrino energy the phase space 
factor reduces the differential rates to zero.) This maximum is broader 
for the plane wave case. Final state interaction effects 
are very important and in the maximum bring the full $d \Gamma_{nd}/dE_{\nu_\mu}$ to about $1/3$ 
of the plane wave prediction. The results are based on the AV18 \cite{av18}
nucleon-nucleon interaction.

\begin{figure}
\includegraphics[width=10cm]{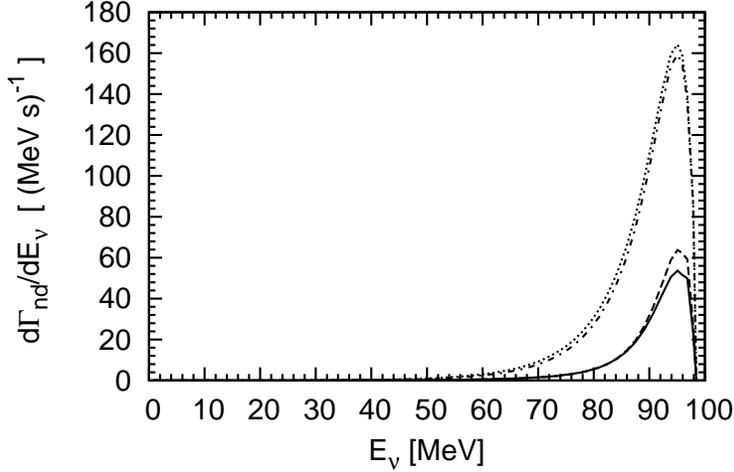}
\caption{The differential capture rates $ {d\Gamma }_{nd}/ {dE_\nu} $
for the $ \mu^- + ^{3}{\rm He} \rightarrow \nu_\mu + n + d$ 
process calculated with the single nucleon current operator and
different types of 3N dynamics: 
plane wave (dash-dotted curve), 
symmetrized plane wave (dotted curve),
full solution of Eq.~(\ref{u}) 
without (dashed curve) 
and with 3N force (solid curve).
The calculations are based on the AV18 nucleon-nucleon potential \cite{av18}
and the Urbana~IX 3N force \cite{urbana} 
and employ all partial wave states with $j \le 3$ and $J \le \frac92$.\label{fig5}}
\end{figure}

\begin{figure}
\includegraphics[width=7cm]{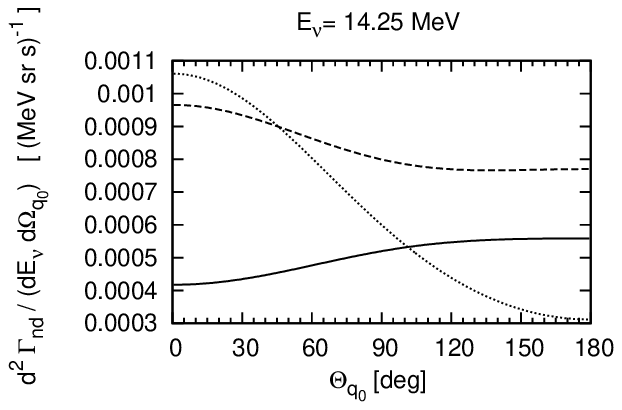}
\includegraphics[width=7cm]{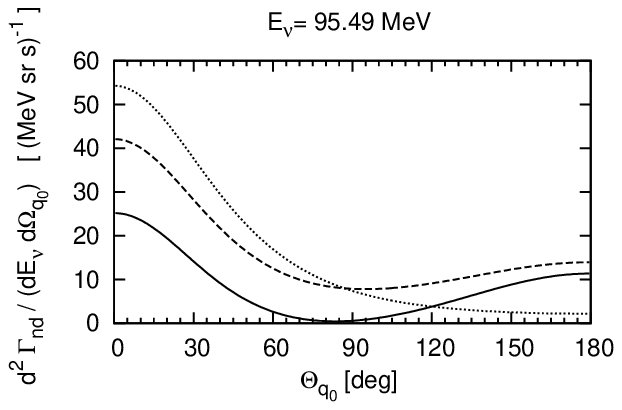}
\caption{The double differential capture rates $ {d^2\Gamma }_{nd}/ ( {dE_\nu} d {\Omega}_{q_0} ) $
for the $ \mu^- + ^{3}{\rm He} \rightarrow \nu_\mu + n + d$ 
process calculated with the single nucleon current operator and using 
the plane wave (dotted curve), 
symmetrized plane wave (dashed curve) and
full solution of Eq.~(\ref{u}) 
but with $ V_4^{(1)} =0 $ (solid curve)
for two values of the neutrino energy.
The calculations are based on the AV18 nucleon-nucleon potential \cite{av18}
and employ all partial wave states with $j \le 3$ and $J \le \frac92$.\label{fig6}}
\end{figure}

In Fig.~\ref{fig5} we show results based on different 3N dynamics: 
plane wave approximation,
symmetrized plane wave approximation,
with the 3N Hamiltonian containing only 2N 
interactions and finally including also a 3N force (here the Urbana~IX 
3N potential \cite{urbana}) 
both in the initial and final state. The effect of the 3N force 
on $ {d\Gamma }_{nd}/ {dE_\nu} $ is clearly visible, since the maximum is
reduced by about $20 \ $\%.
From this figure one might draw the conclusion that the symmetrization
in the plane wave matrix element is not important. We found this agreement
between the plane wave and the symmetrized plane wave results rather 
accidental. As demonstrated in Fig.~\ref{fig6} for two neutrino energies,
the double differential capture rates 
$ {d^2\Gamma }_{nd}/ ( {dE_\nu} d {\Omega}_{q_0} ) $ receive dominant 
contributions from different angular regions.

\begin{figure}
\includegraphics[width=10cm]{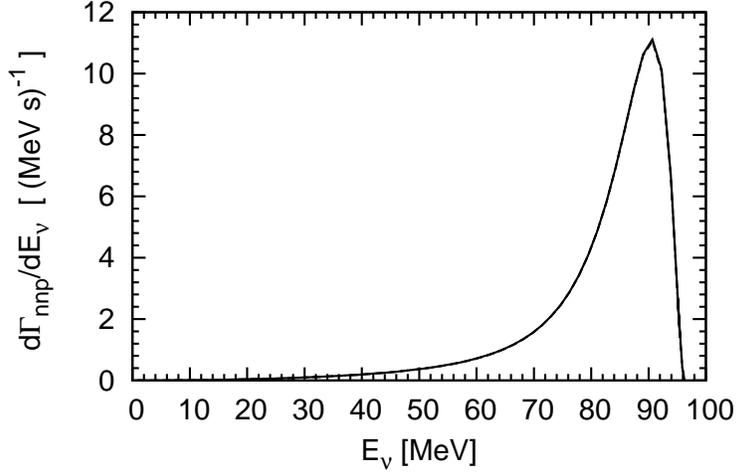}
\caption{The differential capture rates $ {d\Gamma }_{nnp}/ {dE_\nu} $
for the $ \mu^- + ^{3}{\rm He} \rightarrow \nu_\mu + n + n + p$ 
process calculated with the AV18 potential \cite{av18}
and using a full solution of Eq.~(\ref{u}) with $ V_4^{(1)} =0 $.
The curves representing results of the calculations
employing all partial wave states with $j \le 3$ ($j \le 4$) in the 2N subsystem 
are depicted with dashed (solid) curves.
The maximal total 3N angular momentum is $J_{max} = \frac92$. \label{fig7}}
\end{figure}

\begin{figure}
\includegraphics[width=10cm]{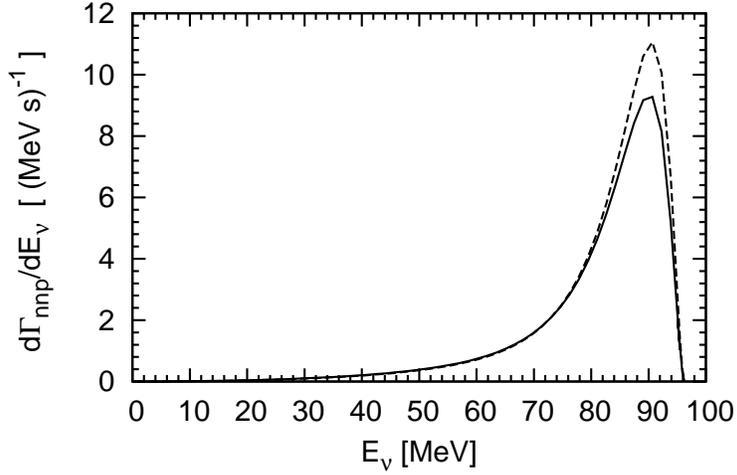}
\caption{The differential capture rates $ {d\Gamma }_{nnp}/ {dE_\nu} $
for the $ \mu^- + ^{3}{\rm He} \rightarrow \nu_\mu + n + n + p$ 
process calculated with full solutions 
of Eq.~(\ref{u}) with $ V_4^{(1)} =0 $ (dashed curve) 
and with $ V_4^{(1)} \ne 0 $ (solid curve).
The calculations are based on the AV18 nucleon-nucleon potential \cite{av18}
and the Urbana~IX 3N force \cite{urbana} 
and employ all partial wave states with $j \le 3$ and $J \le \frac92$.\label{fig8}}
\end{figure}

For the  
$ \mu^- + ^3{\rm He} \rightarrow \nu_\mu + n + n + p$ reaction
we show in Fig.~\ref{fig7} that the convergence 
of the differential capture rate $ {d\Gamma }_{nnp}/ {dE_\nu} $
with respect to the number of partial wave states used in the full
calculations is also very good. Comparing the shapes of
$ {d\Gamma }_{nd}/ {dE_\nu} $ and $ {d\Gamma }_{nnp}/ {dE_\nu} $
we see that the latter becomes significantly different from zero 
at smaller neutrino energies. The calculations are based in this case
on the AV18 \cite{av18} nucleon-nucleon potential and 3N 
force effects are neglected.
In Fig.~\ref{fig8} we show 3N force effects 
adding the Urbana~IX 3N force to the Hamiltonian.
The peak reduction caused by
the 3N force amounts to about $19\ $\%
which is quite similar to the two-body break-up case.
Note that this dependence on the 3N interaction, or
essentially on the trinucleon binding energy, is presumably a consequence
of the overprediction of the $A=3$ radii when 3N interaction
is not included. 

We supplement the results presented in Figs.~\ref{fig4}--\ref{fig8}
by giving the corresponding values of integrated capture rates
in Table~\ref{tabbr}, together with earlier
theoretical predictions of Refs.~\cite{yanox64,phili75,congl94} 
and  experimental data from Refs.~\cite{zaimi63b,auerb65,maevx96,pra69.012712}. 
From inspection of the table we can conclude, first of all, 
that our results are fully at convergence. Secondly, we can estimate 
3N force effects for the total rates.
For the two break-up reactions separately ($\Gamma_{nd}$ and $\Gamma_{nnp}$)
as well as for the total break-up capture rate ($\Gamma_{nd} + \Gamma_{nnp}$)
we see a reduction of their values by 
about $10 \ $\%, when the 3N force is included.
Our best numbers (obtained with the AV18 nucleon-nucleon potential 
and Urbana~IX 3N force and the single nucleon current operator) are  
$\Gamma_{nd} \ = \ 544 \ {\rm s}^{-1} $,
$\Gamma_{nnp} \ = \ 154 \ {\rm s}^{-1} $
and $\Gamma_{nd}+\Gamma_{nnp} \ = \ 698 \ {\rm s}^{-1} $
and can be compared with the available experimental data 
gathered in Table~\ref{tabbr}, finding an overall
nice agreement between theory and experiment for $\Gamma_{nd}+\Gamma_{nnp}$, 
except for the two results of Refs.~\cite{phili75,congl94}. 
The experimental uncertainties are however quite large. 
When comparing with the latest experimental values of 
Ref.~\cite{pra69.012712}, we find that
our results for $\Gamma_{nnp} $ are smaller than the experimental 
values and fall within the experimental estimates 
for $\Gamma_{nd}$ and  $\Gamma_{nd}+\Gamma_{nnp} $.
We expect that our predictions will be changed by about 10~\%, when
many body current operators are included in our framework,
as in the case of 
$\mu^- + ^3{\rm He} \rightarrow \nu_\mu + ^3{\rm H}$.

\subsection{Analysis of the most recent experimental data for the 
differential capture rates}
\label{section6}

Next we embark on an analysis of experimental differential 
capture rates  
$ {d\Gamma }_{nd}/ {dE_d} $
and 
$ {d\Gamma }_{nnp}/ {dE_p} $
published in Ref.~\cite{pra69.012712}.
For a number of deuteron and proton energies 
these quantities are averaged over 1~MeV-wide energy intervals 
and presented in the form of tables.
The tables contain experimental results normalized to 1 in given energy regions 
as well as absolute values.
The data and their uncertainties have been obtained by two different 
methods so in each case two data sets are available.
The first method uses Monte Carlo simulations
and $\chi^2$ minimization procedure to compare simulated 
results, depending on a set of parameters, with experimental events.
In the second approach a Bayesian estimation is used 
to determine the energy distributions
of protons and deuterons emitted in the caption reactions.

One could, in principle, prepare a dedicated kinematics to deal 
with this kind of energy bins, as we did in Ref.~\cite{romek}. 
Our approach is now, however, quite different and very simple.
We have already calculated 
the capture rates $ {d\Gamma }_{nd}/ {dE_\nu} $
and 
$ {d\Gamma }_{nnp}/ {dE_\nu} $
on a dense grid (60 points) of neutrino energies,
solving for each neutrino energy the corresponding 
Faddeev-like equation (\ref{u}).
These neutrino energies are distributed uniformly in the whole 
kinematical region and some extra points are calculated close
to the maximal neutrino energy. 
This dense grid allows us to use the 
formulas and codes which calculate the total 
$ {\Gamma }_{nd}$ (\ref{gnd}) and ${\Gamma }_{nnp}$ (\ref{gnnp}) 
capture rates, performing integrals over the whole phase 
spaces. The sole difference is that in the calculation 
for a given energy interval only contributions 
to the corresponding total capture rate with a proper kinematical ``signature'' 
are summed. 

\begin{figure}
\includegraphics[width=7cm]{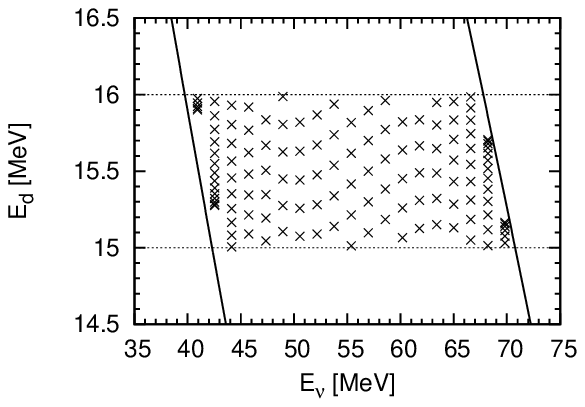}
\includegraphics[width=7cm]{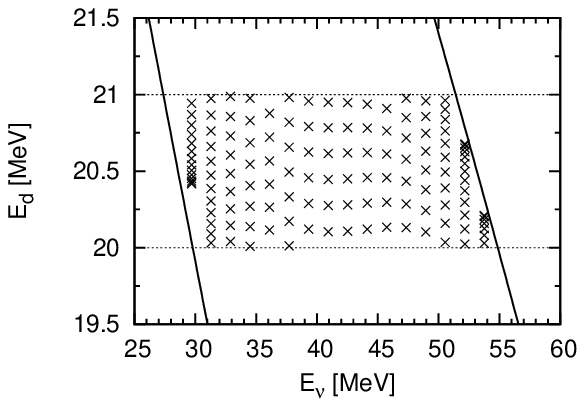}
\caption{The "events" for two selected bins 
corresponding to Fig.~\ref{fig10} 
with the central 
deuteron energy $E_d$= 15.5 MeV (left panel) and 20.5 MeV (right panel),
generated with 60 uniformly distributed $E_\nu$ points 
in the $ [ 0,  E_\nu^{max,nnp} ] $ interval and 72 uniformly distributed $\theta_{q_0}$
values of the relative momentum ${\bf q}_0$ (in $ [ 0, \pi ]$) as explained in the text.
For the selected examples the number of "events" is approximately equal to 130.
\label{fig10.5}}
\end{figure}

This kinematical ``signature'' is easy to obtain.
In the case of the two-body break-up reaction
it is given by Eq.~(\ref{q0}), which can be used to calculate the deuteron
momentum and thus its kinetic energy. 
Two examples showing the distributions of ``events'' 
for two deuteron energy intervals in the $E_\nu - E_d$ plane
are given in Fig.~\ref{fig10.5}. 
The central deuteron energies are 15.5 MeV and 20.5 MeV. 
In this case the events are generated by different 
($E_\nu$, $\theta_{q_0}$) pairs.
 
\begin{figure}
\includegraphics[width=7cm]{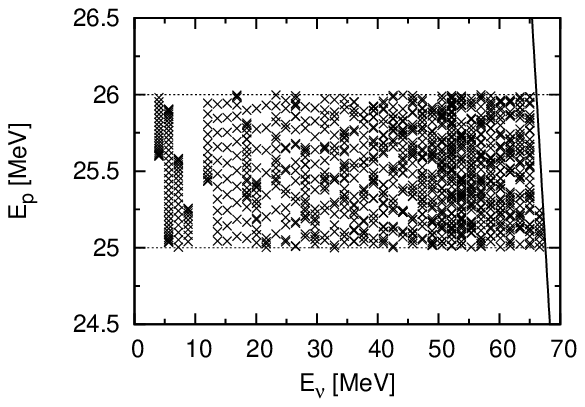}
\includegraphics[width=7cm]{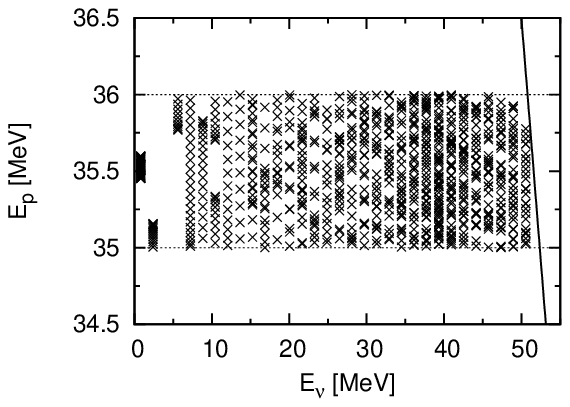}
\caption{The "events" for two selected bins 
corresponding to Fig.~\ref{fig12} 
with the central 
proton energy $E_p$= 25.5 MeV (left panel) and 35.5 MeV (right panel),
generated with 60 uniformly distributed $E_\nu$ points, 
36 uniformly distributed $\theta_{q}$ values 
of the relative momentum ${\bf q}$
and 32 values of the magnitude of ${\bf q}$ (see text for a detailed 
explanation). For these two examples the number of "events" is 
approximately 1000.
\label{fig12.5}}
\end{figure}

For the three-body break-up reaction the proton energy  
can be evaluated from Eqs.~(\ref{pq}). 
Again we demonstrate in Fig.~\ref{fig12.5} 
two examples showing the distributions of proton ``events''
for two proton energy intervals in the $E_\nu - E_p$ plane.
(The central proton energies are 25.5 and 35.5 MeV.)
We see much more events than in the deuteron case, now
generated with 60 uniformly distributed $E_\nu$ points,
36 uniformly distributed $\theta_{q}$ values
of the relative momentum ${\bf q}$
and 32 values of the magnitude of $q \equiv \mid {\bf q} \mid$.
Compared to the deuteron case, the ``events'' come from much broader 
neutrino energy range.

\begin{figure}
\includegraphics[width=7cm]{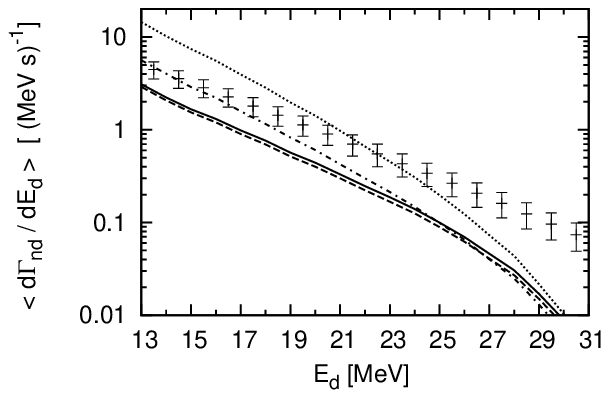}
\includegraphics[width=7cm]{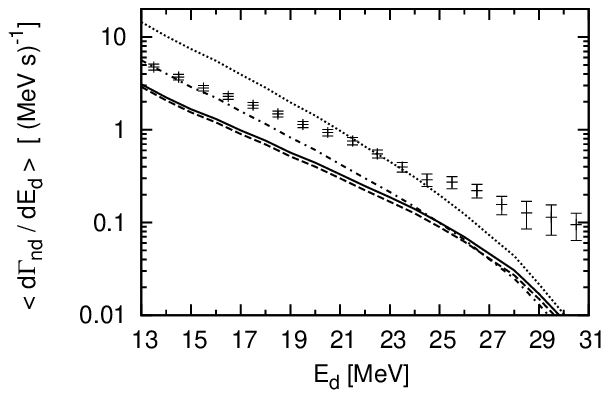}
\includegraphics[width=7cm]{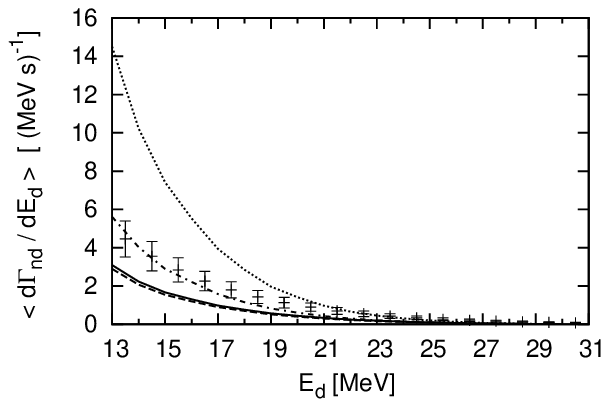}
\includegraphics[width=7cm]{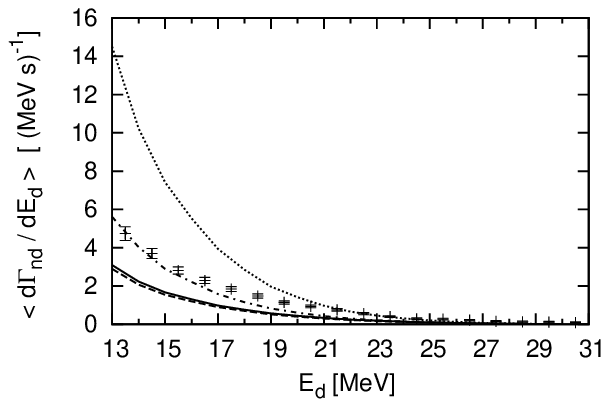}
\caption{The capture rates $ \langle  {d\Gamma }_{nd}/ {dE_d} \rangle $
for the $ \mu^- + ^{3}{\rm He} \rightarrow \nu_\mu + n + d$ 
process averaged over 1~MeV deuteron energy bins are
compared with the experimental data given in Table~VI 
of Ref.~\cite{pra69.012712}.
In the left (right) panel the experimental data are evaluated using 
method~I (method~II) of Ref.~\cite{pra69.012712}.
The notation for the curves is the same of Fig.~\ref{fig5}.
\label{fig10}}
\end{figure}

We show in Fig.~\ref{fig10} 
the capture rates $ \langle  {d\Gamma }_{nd}/ {dE_d} \rangle $
for the $ \mu^- + ^{3}{\rm He} \rightarrow \nu_\mu + n + d$
process averaged over 1~MeV deuteron energy bins, calculated with
various 3N dynamics and compared to the two sets 
of experimental data presented in Table~VI of Ref.~\cite{pra69.012712}.
We show the results both on the logarithmic and linear scales.
Our simplest plane wave calculations  (dash-dotted curves)
describe the data well only for small neutrino energies. Predictions based 
on the full solution of Eq.~(\ref{u}) without (dashed curves) and with (solid curves)
a 3N force clearly underestimate the data by nearly a factor of $2$.
If the Urbana IX 3N force \cite{urbana} is added to the 3N 
Hamiltonian based on the AV18 potential \cite{av18}, the agreement with the data
is slightly improved. The symmetrized plane wave approximation overshoots the data 
for smaller neutrino energies and drops much faster than data at higher 
neutrino energies. 

\begin{figure}
\includegraphics[width=7cm]{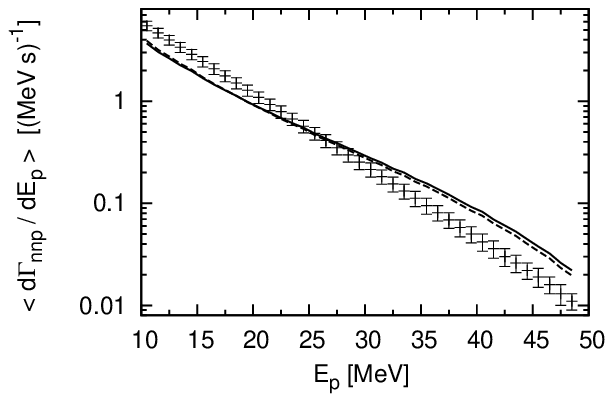}
\includegraphics[width=7cm]{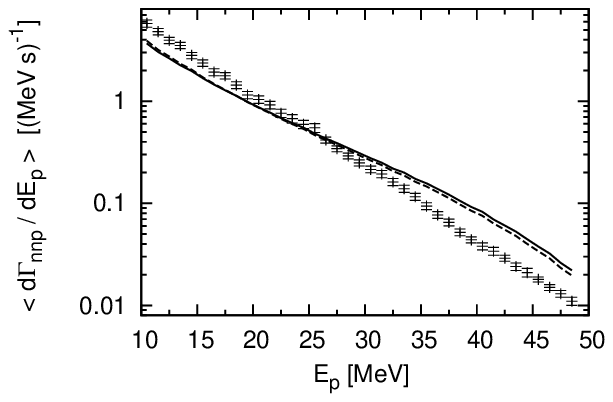}
\includegraphics[width=7cm]{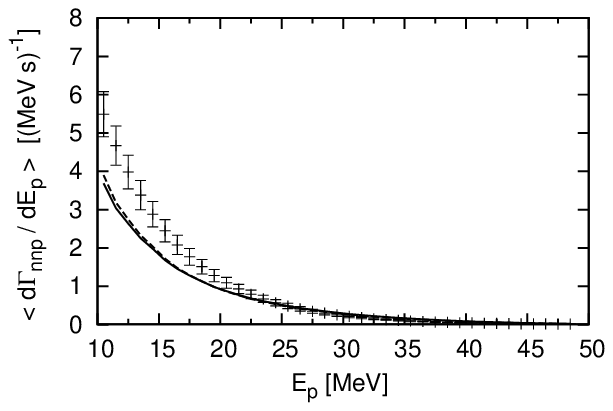}
\includegraphics[width=7cm]{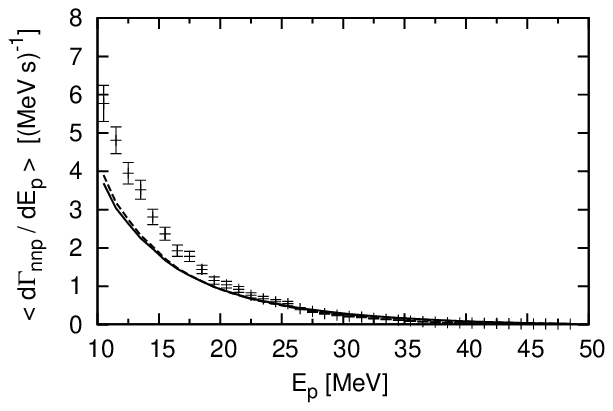}
\caption{The capture rates $ \langle  {d\Gamma }_{nnp}/ {dE_p} \rangle $
for the $ \mu^- + ^{3}{\rm He} \rightarrow \nu_\mu + n + n + p$ 
process averaged over 1~MeV proton energy bins
are compared with the experimental data shown in Table~V 
of Ref.~\cite{pra69.012712}.
In the left (right) panel the experimental data are evaluated using 
method~I (method~II) of Ref.~\cite{pra69.012712}.
The notation for the curves is the same of Fig.~\ref{fig8}.
\label{fig12}}
\end{figure}

The situation for the averaged capture rates 
$ \langle  {d\Gamma }_{nnp}/ {dE_p} \rangle $
in the case of the $ \mu^- + ^{3}{\rm He} \rightarrow \nu_\mu + n + n + p$ reaction
is demonstrated in Fig.~\ref{fig12}. Here we compare 
our predictions obtained with the full solution of Eq.~(\ref{u}) 
without (dashed curve) and with (solid curve)
the Urbana~IX 3N force \cite{urbana} 
to the experimental data evaluated using two methods 
and shown in Table~V of Ref.~\cite{pra69.012712}.
Both types of theoretical results underestimate the data 
for smaller proton energies and lie much higher than the data
for higher proton energies.
The inclusion of the 3N force does not bring the theory 
closer to the data and the 3N force effects 
are quite tiny.

These two comparisons raise the question whether the calculations 
of the total rates $\Gamma_{nd}$ and $\Gamma_{nnp}$  (where we at least roughly
agree with the data)
are consistent with the calculations of the (averaged) 
differential rates
$ \langle  {d\Gamma }_{nd}/ {dE_d} \rangle $
and
$ \langle  {d\Gamma }_{nnp}/ {dE_p} \rangle $
(where we disagree with the data).
We have checked that this is the case, calculating 
$\Gamma_{nd} ( E_\nu < 90\ {\rm MeV}) $  in two ways.
First we used the information given by $ {d\Gamma }_{nd}/ {dE_\nu} $.
In the second calculation we generated corresponding ``events'' 
for all deuteron energies provided that $E_\nu < 90\ {\rm MeV}$
and later used the code for 
$ \langle  {d\Gamma }_{nd}/ {dE_d} \rangle $
to sum the corresponding contributions.

One might also worry if the extrapolation of the experimental results 
(necessary to arrive at the total rates) made by the authors 
of Ref.~\cite{pra69.012712} is justified. From  
Figs. \ref{enu_ed} and \ref{enu_ep} 
it is clear that the data for these two reactions do not 
cover the region of neutrino energies greater than $90\ $ MeV. 
From our calculations we can see that the total capture rates
receive decisive contributions just from this region.
In the two-body break-up case this contribution amounts to nearly $70 \ $\%.
The simple formula used by the authors 
of Ref.~\cite{pra69.012712} to represent the dependence 
$ \langle  {d\Gamma }_{nd}/ {dE_d} \rangle $
on the deuteron energy might not work well for all the deuteron 
energies.
This means that our agreement with experimental 
data for the total rates from Ref.~\cite{pra69.012712} could be more or less 
accidental.
At the moment our theoretical framework is not complete and 
this question should be revisited when the calculations
with the more complete current operator are performed.

\begin{figure}
\includegraphics[width=7cm]{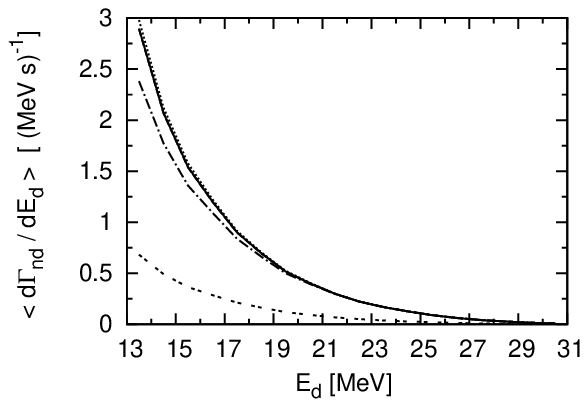}
\includegraphics[width=7cm]{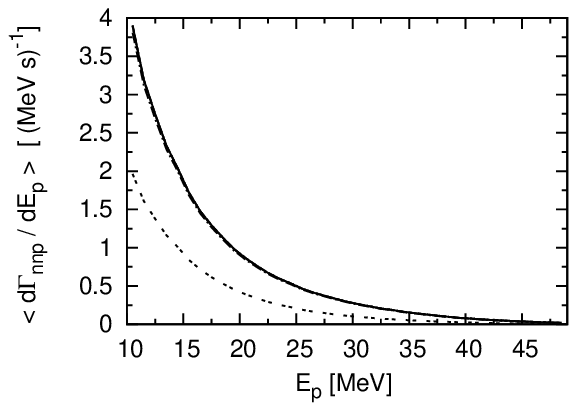}
\caption{Convergence of the full results (without a 3N force) 
with respect to the total angular momentum of the final 3N system 
corresponding to Figs.~\ref{fig10} (left panel) and \ref{fig12} (right panel).
Curves show results of calculations with 
$J \le \frac12 $ (double dashed),
$J \le \frac32 $ (dash-dotted),
$J \le \frac52 $ (dotted),
$J \le \frac72 $ (dashed) and
$J \le \frac92 $ (solid).\label{fig13}}
\end{figure}

Finally, we would like to mention that we used these
more exclusive observables,
$ \langle  {d\Gamma }_{nd}/ {dE_d} \rangle $
and
$ \langle  {d\Gamma }_{nnp}/ {dE_p} \rangle $,
to verify the convergence of the full results 
with respect to the total angular momentum of the final 3N system, $J$.
In Fig.~\ref{fig13} we show
results of calculations performed with 
$J \le \frac12 $,
$J \le \frac32 $,
$J \le \frac52 $,
$J \le \frac72 $,
$J \le \frac92 $.
corresponding to Figs.~\ref{fig10} and \ref{fig12}.
The convergence is extremely rapid, especially in the case
of the 3N break-up reaction and actually 
$J_{max}=\frac92 $ seems unnecessary large.

\begin{table}
\caption{Capture rates for the $\mu^- + ^3{\rm He} \rightarrow \nu_\mu + n + d $
($\Gamma_{nd}$)
and $\mu^- + ^3{\rm He} \rightarrow \nu_\mu + n + n + p $  
($\Gamma_{nnp}$)
processes calculated 
with the AV18~\cite{av18} nucleon-nucleon potential and the 
Urbana~IX~\cite{urbana} 
3N force, using the single nucleon current and describing
the final states just in plane wave (PW), symmetrized plane wave (SPW),
and including final state interaction (full). Early 
theoretial predictions from Refs.~\cite{yanox64,phili75,congl94} are
also shown as well as experimental data are from 
Refs.~\cite{zaimi63b,auerb65,maevx96,pra69.012712}.
\label{tabbr}}
\begin{tabular}{l|c|c|c|c|c}
\hline\hline
& \multicolumn{5}{|c}{capture rate $\Gamma$ in s$^{-1}$} \\ \cline{2-6}
& \multicolumn{3}{|c|}{$\Gamma_{nd}$} & 
$\Gamma_{nnp}$ 
& $\Gamma_{nd}+\Gamma_{nnp}$ \\ \cline{2-6}
& PW & SPW & full & full & full \\ 
\hline
AV18 ($j_{max}= 3$) & 1917 & 2046 & 604 & 169 & 773 \\
AV18 ($j_{max}= 4$) & 1917 & 2046 & 606 & 170 & 776 \\
AV18+Urbana~IX ($j_{max}= 3$) & 1853 & 1956 & 544 & 154 & 698 \\
\hline
earlier theoretical predictions: & \multicolumn{3}{|r|}{}&& \\
A.F. Yano~\cite{yanox64}                           &\multicolumn{3}{|c|}{510}  & 160   & 670 \\
A.C. Philips {\it et al.}~\cite{phili75}           &\multicolumn{3}{|c|}{414}  & 209   & 623 \\
J.G. Congleton~\cite{congl94}         &\multicolumn{3}{|r|}{}&&  650 \\
experimental results: & \multicolumn{3}{|r|}{}&& \\
O.A. Za{\u{\i}}midoroga {\it et~al.}~\cite{zaimi63b}      &\multicolumn{3}{|r|}{}&& $  660 \pm 160 $  \\
L.B. Auerbach {\it et al.}~\cite{auerb65}          &\multicolumn{3}{|r|}{}&& $ 665 \:\:^{+ \: 170}_{- \: 430}$ \\
E.M. Maev {\it et al.}~\cite{maevx96}              &\multicolumn{3}{|r|}{}&& $720 \pm 70 $\\
V.M. Bystritsky {\it et al.}~\cite{pra69.012712}: & \multicolumn{3}{|r|}{}&& \\
\hspace*{0.4cm}method~I      &\multicolumn{3}{|c|}{$491\pm125$}  & $187\pm11$    & $678\pm 126$ \\
\hspace*{0.4cm}method~II     &\multicolumn{3}{|c|}{$497\pm 57$}  & $190\pm7$     & $687\pm60$ \\ 
\hline\hline
\end{tabular}
\end{table}

\subsection{Analysis of the older experimental data for the 
differential capture rates}
\label{section6.5}

In this subsection we provide an analysis of experimental differential 
capture rates  
$ {d\Gamma }_{nnp}/ {dE_p} $
and 
$ {d\Gamma }_{nd}/ {dE_d} $
published in Refs.~\cite{datainromek1,datainromek2}.
For each reaction two data sets were obtained with two different detectors.

\begin{figure}
\includegraphics[width=10cm]{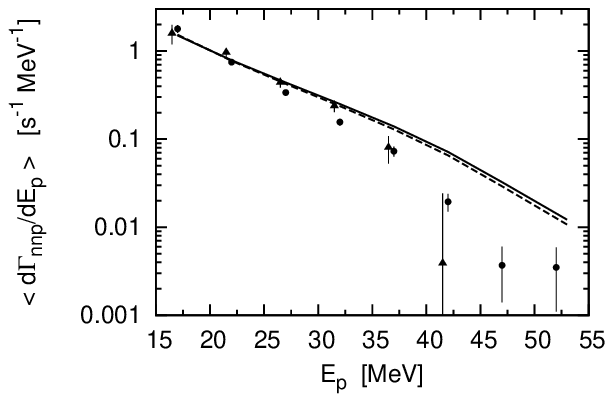}
\caption{The capture rates $ \langle  {d\Gamma }_{nnp}/ {dE_p} \rangle $
for the $ \mu^- + ^{3}{\rm He} \rightarrow \nu_\mu + n + n + p$ 
process averaged over 5~MeV proton energy bins
are compared with the experimental data shown in Table~I
of Ref.~\cite{datainromek2}.
Circles and triangles are used to represent 
data taken at two different detectors.
The notation for the curves is the same of Fig.~\ref{fig8}.
\label{kuhnp}}
\end{figure}

The data for the $ {d\Gamma }_{nnp}/ {dE_p} $ capture rate
are to be found in Table~I of Ref.~\cite{datainromek2}. These data points were averaged over 
5-MeV-wide energy bins and our theoretical predictions are prepared consistently.
The average procedure has been carried out 
in the same way
as described is Sec.~\ref{section6}. The fact that in this case the proton energy
bins are five times larger poses no additional difficulty. We have noticed that 
this additional average over wider proton energy bins does not change significantly 
the representation of our calculations (at least on the logarithmic scale). 
In Fig.~\ref{kuhnp} we see that our calculations are in fair agreement 
with data for $E_p \le 32$ MeV but clearly overshoot the data for the higher
proton energies.

\begin{figure}
\includegraphics[width=10cm]{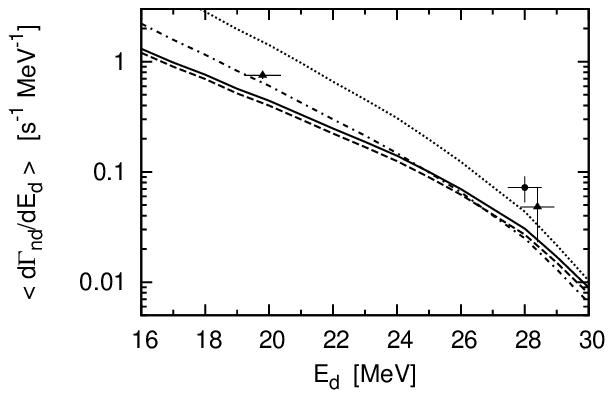}
\caption{The capture rates $ \langle  {d\Gamma }_{nd}/ {dE_d} \rangle $
for the $ \mu^- + ^{3}{\rm He} \rightarrow \nu_\mu + n + d$ 
process averaged over 1~MeV deuteron energy bins are
compared with the experimental data given in Table~III
of Ref.~\cite{datainromek2}.
The notation for the curves is the same of Fig.~\ref{fig5}.
\label{kuhnd}}
\end{figure}

The data set for the $ {d\Gamma }_{nd}/ {dE_d} $ capture rate
consists of three points only. They are given
in Table~III and shown in Fig.~9 of Ref.~\cite{datainromek2}. These data points 
are compared with our 
theoretical predictions (based on different types
of 3N dynamics) averaged over 1-MeV-wide energy bins.
(That means that we use the same results as in the previous subsection.)
This bin width corresponds closely to the horizontal errors bars of the three 
experimental points.
In Fig.~\ref{kuhnd} the simplest plane wave prediction
seems to be consistent with the lower energy datum, while the symmetrized
plane wave result agrees with the higher energy data. 
The full results both neglecting and including 3N force effects
underestimate also the data from Refs.~\cite{datainromek1,datainromek2},
missing them by 40~\% -- 60~\%.
The same data were analyzed by some of the authors of the present paper in Ref.~\cite{romek}
with older nucleon-nucleon forces and without 3N potentials.
Here we do not confirm the results of Ref.~\cite{romek}, which showed a big difference
between the full and symmetrized plane wave predictions.
This might indicate some problems in calculations of Ref.~\cite{romek}
and will be further investigated.

\section{Summary and conclusions}
\label{section7}
A consistent framework 
for the calculations of all muon capture processes on
the deuteron, $^3$He and other light nuclei 
should be ultimately prepared. This requires that the initial and final nuclear states 
are calculated with the same Hamiltonian and that the weak current operator 
is ``compatible'' with the nuclear forces. 
If results of such calculations can be compared with precise 
experimental data, our understanding of muon capture (and other) important weak 
reactions will be definitely improved.

In the present paper we studied
the
$\mu^- + ^2{\rm H} \rightarrow \nu_\mu + n + n$,
$\mu^- + ^3{\rm He} \rightarrow \nu_\mu + ^3{\rm H}$,
$\mu^- + ^3{\rm He} \rightarrow \nu_\mu + n + d$
and
$\mu^- + ^3{\rm He} \rightarrow \nu_\mu + n + n + p$
reactions
in the framework close to the potential model
approach of Ref.~\cite{prc83.014002} but (except for one attempt) 
with the single nucleon current operator.
Contrary to Ref.~\cite{prc83.014002}, we work exclusively in the momentum space.
In all the cases we check carefully that the nonrelativistic 
kinematics can be safely used and outline the adopted approximations.
We also prove the convergence of our results with respect to the number
of partial wave states used in our calculations.

In the case of the $\mu^- + ^2{\rm H} \rightarrow \nu_\mu + n + n$
reaction we employed our scheme, which totally avoids standard partial 
wave decomposition to cross check further elements of our framework.
We supplement information given in the literature by showing some 
predictions for the quadruplet differential and total capture rates.
Already in the 2N system we have developed an easy and 
efficient way to deal with PWD of any single nucleon 
operator. This scheme is then employed also in the reactions with $^3$He.

We give first realistic predictions 
for the differential 
$d\Gamma_{nd} /dE_{\nu_\mu}$ 
and
$d\Gamma_{nnp} /dE_{\nu_\mu}$ 
capture rates as well as for the corresponding total capture rates
$\Gamma_{nd}$ 
and
$\Gamma_{nnp}$. 
Our numbers calculated 
with the AV18 nucleon-nucleon potential \cite{av18}
and the 3N Urbana~IX potential \cite{urbana}
are 544 s$^{-1}$ ($\mu^- + ^3{\rm He} \rightarrow \nu_\mu + n + d$)
and
154 s$^{-1}$ ($\mu^- + ^3{\rm He} \rightarrow \nu_\mu + n + n + p$).

Our analysis of the experimental data from Ref.~\cite{pra69.012712}
reveals some contradictions. We agree roughly with the total 
capture rates but fail to reproduce the differential capture rates.
Our results might indicate that the extrapolations 
and the experimental results on the total 
capture rates published in Ref.~\cite{pra69.012712} should be reconsidered.
Finally, we are well aware that 
the full understanding of the muon capture processes requires
the inclusion of at least 2N 
contributions to the nuclear current operators. However, the work
presented here is a first step to perform a complete
calculation in the near future. Work along this line is currently
underway. Nevertheless, the presented predictions will serve 
as an important benchmark for the future.

\acknowledgments
This study was supported by the Polish National Science Center under Grant No.DEC-
2013/10/M/ST2/00420.
We acknowledge support by the Foundation for Polish Science-MPD program, co-financed by the European Union within the Regional Development Fund.
The numerical calculations have been performed on the supercomputer clusters of the JSC, J\"ulich,
Germany.

\end{document}